\documentclass[11pt]{article}
                
\usepackage{graphicx}   
\usepackage{dcolumn}    
\usepackage{bm}         
\usepackage{amsmath}
\usepackage{amssymb}
\usepackage[usenames]{color}      

\usepackage{tcolorbox}
  
\usepackage{graphicx}
\usepackage{subfigure,comment}
\usepackage{epsfig}
\usepackage{color}
\usepackage{fancyhdr}
\usepackage{citesort}
\usepackage{url}
  
\usepackage{caption}
\usepackage{wrapfig}
\usepackage{sidecap}
\usepackage{multirow}
\usepackage{colortbl}
\usepackage{textgreek}
\usepackage{array}

\usepackage{amsmath}
\usepackage{amssymb}
\usepackage{wasysym}
\usepackage{epstopdf}
\usepackage{xcolor}
\usepackage{epstopdf}
\usepackage{arydshln}

\newcommand\euro{{\sffamily C%
    \makebox[0pt][l]{\kern-.70em\mbox{--}}%
    \makebox[0pt][l]{\kern-.68em\raisebox{.25ex}{--}}}}

\def\be{\begin{equation}}
\def\ee{\end{equation}}

\newcommand{\bea}{\begin{eqnarray}}
\newcommand{\eea}{\end{eqnarray}}

\def\x{{\mbox{\boldmath$x$}}}

\def\v{{\mbox{\boldmath$v$}}}

\def\begineq{\begin{equation}}
\def\endeq{\end{equation}}

\begin{document}

\title{\bf
Physicochemical Hydrodynamics of 
Droplets 
\\ out of  Equilibrium:  A Perspective Review 
}

\author{Detlef Lohse\footnote{d.lohse@utwente.nl, Physics of Fluids Group, 
Max-Planck Center for Complex Fluid Dynamics, MESA+ Research Institute and J.M. Burgers Centre for Fluid Dynamics, University of Twente, P.O. Box 217, 7500 AE Enschede, The Netherlands, and Max Planck Institute for Dynamics and Self-Organization, Am Fassberg 17, 37077 G\"ottingen, Germany}   ~and Xuehua Zhang\footnote{xuehua.zhang@ualberta.ca, Department of Chemical and Materials Engineering, University of Alberta, Edmonton, AB T6G 1H9,  Canada, and  Physics of Fluids Group, Max-Planck Center for Complex Fluid Dynamics, MESA+ Research Institute and J.M. Burgers Centre for Fluid Dynamics, University of Twente, P.O. Box 217, 7500 AE Enschede, The Netherlands.}}



\maketitle


\begin{abstract}
Droplets  abound 
 in nature and technology. In general,  they are 
{\it multicomponent}, and, when out of equilibrium, 
with gradients in concentration, implying  flow and mass transport. 
Moreover, {\it phase transitions} can 
occur, with either evaporation, solidification, 
dissolution, or nucleation of a  new phase. The droplets and their surrounding 
 liquid  can be binary, ternary, or contain even more 
components, and with several even  in different phases.
In the last two decades the rapid advances
 in experimental and numerical  fluid dynamical techniques 
has enabled major progress in our understanding of the physicochemical hydrodynamics of such droplets, 
further narrowing  the
gap from 
fluid dynamics  to chemical engineering and colloid \& interfacial science and arriving at a  quantitative  understanding 
of multicomponent and multiphase droplet  systems far from equilibrium, and aiming towards 
a one-to-one comparison 
between experiments and theory or numerics. 

This review will discuss various  examples of the physicochemical hydrodynamics of droplet systems far from 
equilibrium and our  present understanding of them. 
These include immiscible droplets in a concentration gradient, coalescence of droplets of different liquids,
droplets in concentration gradients emerging from chemical reactions (including droplets as  microswimmers)
and phase transitions such as evaporation, solidification, dissolution,
 or nucleation,  and droplets in ternary liquids, including 
solvent exchange, nano-precipitation,  and the so-called ouzo effect.
We will also discuss the relevance 
 of the physicochemical hydrodynamics of such droplet systems for many important applications, 
 including in chemical analysis and diagnostics,  microanalysis,  pharmaceutics,
 synthetic chemistry and biology, 
 chemical and environmental engineering, the oil and remediation industries, 
 inkjet-printing,  for micro- and nano-materials, and in nanotechnolgoy. 
\end{abstract}

\maketitle

\newpage

\section{Introduction}\label{intro}
Classical hydrodynamics focuses on pure liquids. In nature and technology, 
fluid dynamical systems are however
{\it multicomponent}, and often with gradients in concentration, even changing in time, i.e., they are out of equilibrium.
These concentration gradients, be they smooth   or sharp,  
can induce a flow. Moreover, phase transitions can 
occur, with either evaporation, solidification, 
dissolution, or nucleation of a  new phase. The liquids can be binary, ternary, or contain even more 
components,  with several across  different phases. The non-equilibrium can be driven by flow, mixing, 
phase transitions, chemical 
reactions, electrical current, heat, etc.

To theoretically deal with such systems, in the 1950s, Levich wrote  the wonderful book 
``Physico\-chemical Hydrodynamics'' \cite{levich1962}. 
The central theme of the book is the ``elucidation of mechanisms of transport
phenomena and the conversion of understanding so gained into plain, useful tools for applications,'' as L.\ E.\ Scriven
describes it  in the foreword to the book in its 1962 translation into English \cite{levich1962}.   
Levich can be viewed  as a physical chemist and theoretical physicist at the same time, or, as Scriven puts it,
 above all,
as an ``engineering scientist'' with a recognizable  ``blend of applied chemistry, applied physics, and fluid mechanics."
V.\ Levich himself, in his foreword to the first Russian Edition (1952), describes the scope of physicochemical hydrodynamics as 
the ``aggregate of problems dealing with the effect of fluid flow on chemicals or physicochemical transformations as well as the effect of physicochemical factors of the flow.''

First and foremost, Levich's ``Physicochemical hydrodynamics''  
 is a book describing the relevant  theoretical and mathematical
concepts, given that in 
those days the experimental tools to actually measure the flow on the microscale were 
 very limited
and the possibility  to perform direct numerical simulations of the underlying partial differential equations 
even absent. 
Today, more than 60 years later, the scientific and in particular the hydrodynamical and physicochemical communities have
developed tremendously and  
the experimental, instrumental, 
 and numerical  means to actually deal with the problems Levich defined in his book have become 
available and are being used to do so.

These developments are more than timely, as the  relevance 
of physicochemical hydrodynamics of multicomponent and multiphase liquids is 
ever increasing, in order to address the  challenges of mankind for the 21st century. 
These challenges include  energy, namely  storage and batteries,  hydrogen production by electrolysis,
CO$_2$ capture, polymeric solar cell manufactering,
biofuel production,  and catalysis. They also include health and medical issues like chemical analysis and 
diagnostics or the production and purification of drugs, advanced material manufacturing,
environmental issues like flotation, water cleaning, membrane
management,  and 
separation technology, or 
food processing and food safety issues. They also include issues in modern production technologies
such as  additive manufacturing on ever decreasing length scales and inkjet printing, and in  the paint and 
coating industry. 

These challenges  have often been  
approached with a pure engineering approach, and less in the spirit 
of Levich as an engineering scientist. On the other hand, as said above, classical hydrodynamics has focused
on pure and single-phase liquids. 
In the last two decades, the advent of new experimental and numerical tools 
has allowed for a more integrated understanding  of physico\-chemical hydrodynamics 
and to further 
narrow  the gap between classical
  hydrodynamics and  chemical engineering and colloid  \& interfacial science. 
The objective of this  effort  is  to 
improve the  {\it quantitative}  understanding 
of multicomponent and multiphase fluid dynamic
 systems far from equilibrium, in order to master and better control them. 
To achieve this objective, one has  to perform 
 controlled experiments and numerical simulations
for idealized setups, allowing for a {\it one-to-one comparison} between experiments and numerics/theory, 
in order to test the theoretical understanding.
This effort indeed is in 
 the spirit of Levich's  ``Physicochemical  Hydrodynamics'', but now building on and benefiting from
the developments of modern microfluidics, microfabrication, 
digital (high-speed) imaging technology, confocal microscopy, atomic force microscopy,
and various computational techniques
and opportunities  for high-performance computing. These are, in a nutshell, the blessings from what 
can be considered as the golden age of fluid dynamics, which builds on the digital revolution, both on the
experimental and the numerical side. 
Given these developments, and given the necessity in chemical engineering to move towards higher precision
and enhanced control, this effort is indeed very timely.

There is a large number  of
 physical phenomena and effects which come into play  in multicomponent and multiphase 
  liquids far from equilibrium. These include gradients in  concentration, either in the bulk of the liquid
  or on the surface, leading to diffusiophoresis and solutal  Marangoni flow. 
  They include (selective) dissolution 
  of (multicomponent) droplets and bubbles  in host liquids or vice versa their nucleation and growth. They also 
  include the coalescence of droplets consisting of different liquids, possibly 
  with chemical reactions and/or solidification and other transitions from one phase to another. The material parameters which become important are
  the various diffusivities and viscosities 
  of the liquids, their surface tensions and how they depend on the concentrations, 
  the volatilities and mutual solubilities, latent heats, reaction rates, etc.

Obviously, the field of multicomponent and multiphase hydrodynamics is too large to give an
exhaustive review. So we restrict ourselves to 
 small length scales, focusing on the physico\-chemical
hydrodynamics of (multicomponent) 
droplets  far from equilibrium. 
The objective of this perspective review is to show examples of such systems
for which a successful quantitative description and one-to-one comparison between well-controlled 
table-top experiments and theory and numerics 
has been achieved, to identify the complex interplay of the 
 underlying principles,  
 and to put these examples into the context of 
relevant applications of multicomponent and multiphase liquid systems. In particular, we want to
identify the new directions of physicochemical hydrodynamics and want to outline the scientific challenges
and their connections with the technological challenges. 

The core of  the review is section \ref{examples}. In the five   subsections, we give  general  examples for the 
physicochemical
hydrodynamics of (multicomponent) 
droplets  far from equilibrium.
In   the five    subsections of section \ref{relevance} 
 we give application fields and examples for the  relevance 
 of physicochemical hydrodynamics 
 in technology.  Finally, 
in section \ref{conclusions} we will  try to identify the general lessons one learns from these examples and applications 
 and will give an  outlook to further directions and perspectives. For better readability, 
  so that in the main sections we can better focus on the fluid dynamics and the physics,
 we have included three text boxes, namely one on the relevant dimensionless numbers in
 physicochemical hydrodynamics, one for the new experimental methods which have 
 enabled the recent progress, and one
 on the new or improved  numerical methods.

 This  perspective
 review convey ideas, rather than delve into detailed technical aspects or be encyclopedic.
 For these aspects 
 we refer to
 specialized review articles on certain aspects, such as the one 
 by Cates and Tjhung on binary fluid mixtures \cite{cates2018},
 by Lauga and Powers on swimmers \cite{lauga2009}, 
 by Maass {\it et al.} on swimming droplets  \cite{maass2016}, 
 by Moran and Posner and by Golestanian on phoretic self-propulsion \cite{moran2017,golestanian2020},
 by Manikantan and Squires on surfactant dynamics \cite{manikantan2020}, 
 by Cazabat \&  Gu\'ena \cite{cazabat2010} and Erbil \cite{erbil2012} on evaporation of pure liquid sessile droplets,  
 by Sefiane on patterns from drying drops \cite{sefiane2014}, 
 by Lohse and Zhang on surface nanobubbles and nanodroplets \cite{lohse2015rmp}, 
 by Jain on single-drop microextraction \cite{jain2011}, 
 by de Wit on chemo-hydrodynamic patterns and instabilities \cite{wit2020}, 
 etc. 


{\small 
\begin{tcolorbox}
\noindent
{\bf Box 1: Dimensionless numbers for droplets in physicochemical hydrodynamics}\\
In fluid dynamics, it is common to express the ratio of the various forces (or time or length scales) in 
terms of dimensionless numbers. The most famous one is the 
\begin{itemize}
\item
{\bf Reynolds number $Re = U R/\nu$}, which expresses the ratio of inertial  forces to viscous forces. 
In the context of droplets, $R$ is the  droplet radius, 
$U$  the (relative) droplet velocity,  and $\nu$ the kinematic viscosity of the continuous phase. 
\end{itemize} 
In the context of the physicochemical hydrodynamics of droplets, which dissolve or
grow, the perhaps even more relevant dimensionless parameter is the 
\begin{itemize}
\item (diffusive) {\bf Peclet number $Pe = U R/D$}, which compares the advective
 and diffusive time scales,
where $D$ is the mass diffusivity, which in general is much smaller than the viscosity $\nu$, reflecting
the very slow process of molecular diffusion in liquids. 
\end{itemize}
The ratio between viscosity and diffusivity is the 
\begin{itemize}
\item
{\bf Schmidt number $Sc = \nu / D$}, which correspondingly in physicochemical hydrodynamics
is quite large, namely $\sim 10^3$, which on the one hand is the reason for many peculiarities
in physicochemical hydrodynamics, and on the other hand introduces many experimental and 
numerical difficulties. 
\end{itemize}
Once also  thermal effects come into play, the thermal diffusivity $\kappa$ will play a role.
Its ratio to the molecular diffusivity and the kinematic viscosity is expressed as
\begin{itemize}
\item
{\bf Lewis number $Le = \kappa / D$} and {\bf Prandtl number $Pr = \nu/\kappa$}, which 
are typically $\sim 400$ and  $\sim 4$, respectively, 
reflecting that the thermal diffusion  is much faster
than the molecular one, and that for standard liquids thermal  transport is slightly slower than
viscous transport.
\end{itemize}
In hydrochemical hydrodynamics, surface tension effects are crucial. They 
can be expressed either in terms of the 
\begin{itemize}
\item
{\bf Weber number $We = \rho U^2 R/\sigma$}, which is the ratio of inertia to capillarity, where $\sigma$ is the surface tension and $\rho$ the density of the liquid, or the 
\item
{\bf Capillary number $Ca = \eta U/\sigma$}, ratio of viscous  to capillary forces, or the 
\item {\bf Ohnesorge number $Oh= \eta /\sqrt{\rho_w \sigma R} = We^{1/2} / Re$}, ratio of time of viscous damping  to time  of the capillary oscillations, or the 
\item 
{\bf Bond number $Bo = \rho gR^2/\sigma$},  ratio of gravity to capillary forces. 
\end{itemize}

The full richness of hydrochemical droplet fluid dynamics 
only enters once {\it several} liquids and gases come into play, with different surface tensions, 
densities, etc. Then the gradient of the hydrodynamic forces along or across a droplets 
introduces net forces. The most relevant may be the
\begin{itemize}
\item
{\bf Marangoni number $Ma=  R \Delta \sigma / (\rho\nu  D )  $}, which compares the Marangoni forces with 
stabilizing viscous forces and stabilizing mass diffusion. The difference $\Delta\sigma$  in surface tension 
may be due to differences $\Delta c$ in the composition of the liquid,   where $c(\x , t)$ is the concentration field, or due 
to differences $\Delta T$ in the temperature of the liquid.  Roughly, $\Delta \sigma \approx \partial \sigma/\partial c ~ \Delta c$,
and similarly for the temperature, but note that in general the dependence  $\sigma (c)$ is very nonlinear \cite{vazquez1995}. In general, of course, $\sigma (T,c)$, and $\Delta \sigma \approx
\partial_T \sigma |_c \Delta T + \partial_c \sigma |_T \Delta c$.
\end{itemize}
Another way for the system to get out of equilibrium are density differences $\Delta \rho$ to which the gravitational 
acceleration $g$ couples, implying a 
\begin{itemize}
\item 
{\bf Rayleigh number $Ra = g R^3 \Delta \rho / (\rho \nu D)$}, which compares destabilizing buoyancy with stabilizing 
viscosity and diffusivity. 
\end{itemize}
Once chemical reaction come into play, we will also need the 
\begin{itemize}
\item {\bf Damk\"ohler number Da}, which expresses the ratio between  chemical reaction rate and 
(diffusive or convective) mass transport
rate.
\end{itemize}
We will encounter 
 all these numbers in the discussions of the various physicochemical hydrodynamical
effects featured  in this review article. 
\end{tcolorbox}}

\begin{tcolorbox}
{\bf Box 2: Recent progress in experimental  tools to address  physicochemical hydrodynamics}\\
A whole plethora of new and improved experimental techniques in fluid dynamics has enabled the recent
progress in understanding the physicochemical hydrodynamics of droplets and bubbles
far from equilibrium. The most relevant of these techniques are: 
\begin{itemize}
\item  
{\bf Optical visualization}   with optical wavelength resolution and beyond: 
These  techniques include standard, fluorescence,  and confocal microscopy \cite{hell2009,webb1996,tan2016}.
The latter  allows for 3D visualizations, 
but the choice of proper dyes is essential, as adding dyes and/or fluorescent molecules (which are
often  surface active substances) may cause contamination to the system and artefacts.
\item {\bf Digital Holographic Microscopy (DHM)}  can be seen as complementary to confocal microscopy, namely focusing on {\it interfaces} rather than on the 
bulk as standard 
confocal microscopy. The technique -- developed only about a decade ago and up to now mainly used in the biological context 
\cite{marquet2005,garcia2006,merola2011} -- 
is crucial and ideal for  high-precision measurements and 
visualizations of growing and  shrinking droplets with small contact angle, having
a sub-nanometer {\it in-depth} resolution.
\item {\bf Micro-particle-image-velocimetry ($\mu$-PIV)} \cite{wereley2010} allows to 
 obtain information on the velocity fields in and around droplets, even if they are moving, growing, or shrinking.
 \item  {\bf High-speed imaging} with frame rates up to several  million frames per second \cite{versluis2012} and 
 {\bf stroboscopic illumination} by nanosecond laser pulses \cite{bos2011,bos2014}
 allow for excellent temporal resolution. 
 These techniques  also include fluorescent
high-speed imaging  \cite{versluis2012}. 
\item
{\bf Atomic force microscopy (AFM)} allows for nanometer or even atomic resolution both laterally and 
in-depth. In the last two decades, it has developed such that it can now routinely also be used in the liquid phase \cite{lohse2015rmp}. 
With techniques like 
 shock-freezing  \cite{switkes2004} or instantaneous polymerization with the help
of UV radiation \cite{zhang2012softmatter}
 the nucleation and growth process of droplets can be terminated. 
 The latter requires the use of UV polymerisable liquids as solutes. 
After  solidification of the
sessile droplet phase through  UV radiation, 
the other liquid phase can be removed, paving the way for AFM of a solid phase in air, which is much easier than of a liquid phase in another liquid. 
\item {\bf Surface-enhanced Raman spectroscopy (SERS)} \cite{schlucker2014} allows to follow the chemical composition of sessile droplets or bubbles in time. 
\item
{\bf Microfluidics:} Well-defined channels and substrates with designed structures and chemical patterns are possible by microfabrication techniques  \cite{whitesides2006,squires2005,garstecki2006,eijkel2005,mello2006,lach2016}
to  study the growth and collective interactions of bubbles and droplets. 
\end{itemize}
\end{tcolorbox}

\begin{tcolorbox}
{\bf Box 3: Recent progress in numerical tools to address  physicochemical hydrodynamics}\\
Alongside 
the  progress in the experimental techniques (see Box 2), also the progress on the numerical side
was crucial in pushing the field ahead. Next to the ever enhancing pure computational power, also the 
development of new numerical methods and open-source codes and tools contributed to this  progress, which
now often make a one-to-one comparison between experiment and numerical simulations possible. 

The main challenge for the numerical simulations in physicochemical hydrodynamics are the vastly different
time scales for the momentum transfer and for the mass transfer, reflected in the large Schmidt number $Sc\approx
10^3$. This implies that the concentration field must live on a smaller grid than the velocity field and that the 
stepping time scale is extremely small. 
Here, multigrid resolution techniques are a possible way out 
\cite{spandan2018}. These can be used for various numerical schemes. 

The most
relevant techniques for the numerical simulations in physicochemical hydrodynamics are: 
\begin{itemize}
\item
{\bf Finite  element methods:} 
Moving droplets or droplets with mass-exchange 
require a well-resolved interface with the surrounding phases. 
This can be achieved with a  sharp-interface finite element method, where the mesh is always conforming with 
the interface \cite{diddens2017num}. Since in general the droplet moves, this requires to co-move all mesh nodes accordingly with the interface. To that end, 
with  Eulerian-Lagrangian methods originally developed for fluid-structure interaction
\cite{heil2006,heil2011},  
the 
 mesh  can be treated as pseudo-elastic body, 
so that the bulk nodes follow the motion of the interfacial nodes, which is connected via Lagrange multipliers to the kinematic boundary condition of the interface \cite{diddens2017num}. 
This technique has proven to be extremely versatile for binary and ternary droplets 
\cite{tan2016,diddens2017,li2019prl-yanshen,li2019-yaxing}, but the challenge remains to efficiently
 parallelize such codes. 
\item
{\bf 
Finite Difference (FD) with Immersed Boundary methods (IBM)} \cite{seo2011,spandan2018} are an alternative,
allowing for massive parallelization and the calculation of tens of dissolving or growing droplets over a long time
\cite{bao2018,zhu2018}. Just as in finite element methods, a weak point of such simulations is that they require
{\it a-priori}
input in what mode  (e.g.\ constant contact angle (CA-mode) vs constant contact radius (CR-mode) 
\cite{cazabat2010,lohse2015rmp}) droplets or bubbles grow or shrink. 
\item {\bf Phase field methods and Cahn-Hilliard type approaches}  \cite{kim2012pf} employ a diffuse interface,
but easily allow to describe spontaneous phase-separation and nucleation, such as occurring in the ouzo effect \cite{moerman2018}. 
\item   
{\bf Level-set methods} \cite{sussman1994,sethian2003} 
have the advantage that the interfaces between the 
different phases are sharp. They have been extended to include phase changes and interfacial flows 
driven by surface tension gradients \cite{langavant2017,gibou2018}.
\item 
{\bf Lattice Boltzmann (LB) methods} \cite{chen1998}  
-- coarse-grained versions of the molecular theory of fluids enabling massive parallelization --
have been adopted and applied to 
multiphase flows  \cite{aidun2010,perlekar2014}, including the selective 
evaporation or dissolution of multicomponent droplets \cite{hessling2017}.
\item 
{\bf Molecular Dynamics (MD) simulations} \cite{koplik1995arfm,frenkel1996,lauga2007springer,bocquet2010}
have the advantage that they do not require to predetermine whether the (sessile) droplet grows  or shrinks 
 in the CA or CR mode as in FD+IBM, but the mode will come up from the respective chemical properties of the substrate and the liquids involved, see e.g.\ refs.\ \cite{maheshwari2016b,maheshwari2018}. 
 The downsides of MD are  that (i) only small nanodroplets ($\sim$50 nm) can be simulated,
 that (ii) one is restricted to short physical times of at most microseconds, 
 and that (iii) molecular interaction potentials are needed, which are not from first principles. 
\end{itemize}
\end{tcolorbox}

\section{Recent fundamental research on physicochemical hydrodynamics in droplet systems}\label{examples}

In this section we will give examples 
for the physicochemical hydrodynamics in selected simple droplet  systems. 
We will focus on systems in which 
modern experimental and numerical techniques 
have led to considerable progress in our understanding, 
often even allowing for a one-to-one comparison between
experiments and numerics. 
These modern techniques are highlighted
in boxes 2 (experimental) and 3 (computational). 
In the next section we will focus on 
the applications and the relevance of these  examples.
The focus is on droplets -- 
for a very general overview on fundamental and applied
 aspect of bubble systems, we refer to 
 our recent review \cite{lohse2018-bp}.

\begin{figure}[htb]
\begin{center}
\includegraphics[width=0.92\textwidth]{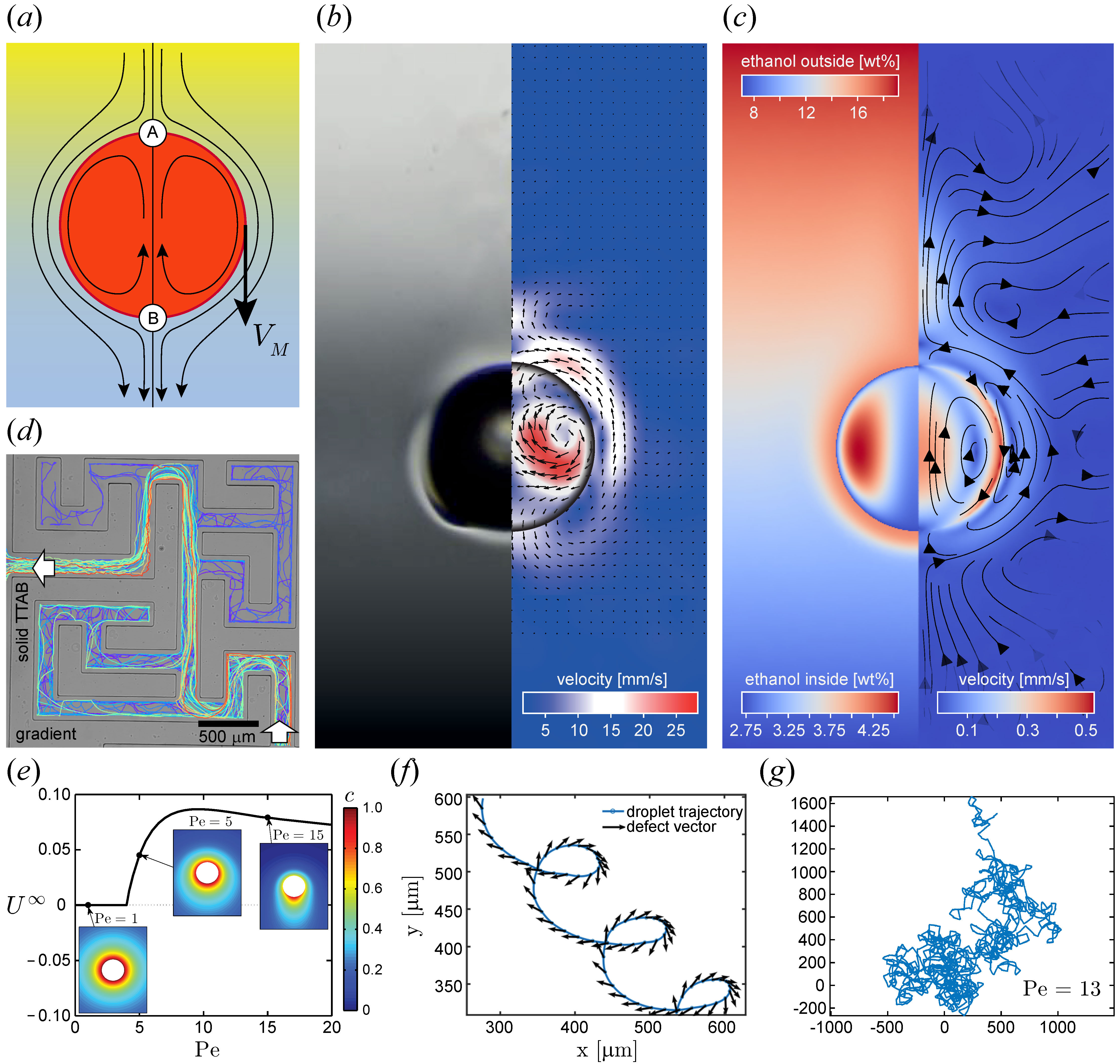} 
\caption{{  \it 
(a) Immiscible droplet (red) in a concentration gradient, shown through the color gradient, with 
the heavier water in the bottom (bluish)
and the lighter ethanol at the top (yellowish). We will stick to this color code in the sketches throughout the article.  
The surface tension at the top of the drop (A) is smaller  than at the bottom (B), leading to some 
Marangoni flow, whose streamlines (calculated in the viscous limit, assuming constant density \cite{young1959}) 
are shown (in the frame of reference of the drop). Also shown is the Marangoni velocity $V_M$. 
The velocity fields  for a 
more complicated configuration of a partially (but poorly) miscible  anise oil drop in 
an ethanol-water concentration gradient are shown in (b) (experimental, from a snapshot and from
PIV images) and (c) (numerical result). 
On the right of each of these figures, the  
velocity field is shown,  both as color code and as stream lines (in the lab frame). 
On the left, in (b) the snapshot is shown and in (c) 
the ethanol concentration inside and outside the anise oil  drop (also shown in the lab frame).
Figures (b) and (c) 
refer to the situation of  ref.\ 
\cite{li2019prl-yanshen}. 
(d) Maze solving by chemotactic droplet swimmers: A chemical gradient of the 
 ionic surfactant tetradecyltrimethylammonium bromide (TTAB) was imposed; 
 the oil droplet propelling in the gradient inside the maze  consists of the nematic liquid crystal 
 4-pentyl-4'-cyano-biphenyl.
The trajectories of the swimmers that passed both entrance and exit points are shown, with the  line color
 corresponding  to the time in the experiment. Figure taken from ref.\ \cite{jin2017pnas}.
 (e)
Long-time spontaneous swimming velocity $U^\infty$ as function of the Peclet number $Pe$.
The concentration field around the droplet (or particle) is shown for three different Peclet numbers. 
A clear bifurcation towards droplet motion thanks to spontaneous symmetry breaking is seen at $Pe=4$.
Figure adapted  from ref.\ \cite{michelin2013}.
(f) Curling nematic 
droplet ($R \approx 25 \mu m$, phoretic)
from the experiments of ref.\ \cite{krueger2016}, where the Peclet number is
estimated to be larger than 4. 
(g) Chaotic droplet motion for a phoretic droplet with an even larger Peclet number $Pe=13$. 
Figure taken from the numerical simulations of \cite{hu2019}.
    }}
\label{young-diddens}
\end{center}
\end{figure}

\subsection{Immiscible droplet in a concentration gradient} \label{sec-yanshen} 
One of the most basic examples  for a droplet far from physicochemical equilibrium is the one of 
an immiscible  (oil) droplet (of radius $R$) in a concentration gradient of two other miscible liquids (with density 
lighter respectively heavier than the oil droplet), see figure \ref{young-diddens}a. 
The two control parameters reflecting the non-equilibrium situation are the Marangoni number $Ma$
defined with the gradient of the surface tension, i.e., $\Delta \sigma =  R \partial_z \sigma $,
 and the Rayleigh number $Ra$, defined with the gradient in density, $\Delta \rho =  R \partial_z \rho $;
 see box 1 for the proper nondimensionalization. 
For the case of Stokes flow (small Reynolds number) and zero density gradient, 
 the resulting velocity and the concentration fields can be analytically calculated \cite{young1959}. 
 The former can be quantified by the 
 Marangoni velocity $V_M$ at the equator  of the 
 droplet or, in dimensionless form, the {{(Marangoni-)Peclet}} number $Pe_{M} = V_M  R/D$. 
 If large enough, i.e., for a large enough concentration gradient, this Marangoni flow can 
 make the droplet jump upwards in the stratified density gradient, against gravity \cite{li2019prl-yanshen}!

In more detail, what happens is the following series of events, that we will describe 
for  the concrete example of 
an anise  oil droplet in a stratified (fully miscible)
mixture of water (bottom) and ethanol (top) with constant
density gradient in between, see figure \ref{young-diddens}a,b,c and ref.\ \cite{li2019prl-yanshen}: 
First, the oil droplet sinks in the gradient region, trying to find its equilibrium position with 
respect to density. Ethanol-rich liquid is entrained downwards during this motion. This entrainment
is enhanced by the Marangoni flow along the drop surface, from the ethanol-rich top (low surface 
tension) to the ethanol-poor bottom, leading to further entrainment of ethanol-rich liquid.
As the droplet asymptotically approaches the density matched position, the 
self-enhancing Marangoni flow eventually overcomes the sinking-induced buoyancy jet.  
This positive feed-back leads to a linear instability and 
exponential growth of the Marangoni flow, which pushes the liquid around the drop downwards
and thus the drop upwards, like a micro-swimmer in the pulling mode \cite{lauga2009}. Once this Marangoni flow is dominant (typically after $\sim 60s$), the drop shoots upwards
towards the low density region and the process can start over. Up to six hours of  consecutive jumps have been observed. 
The process only comes to an end once the jumping droplet has sufficiently mixed
the stratified liquid. 
 The system can be seen as canonical example for the competition between
Marangoni (surface tension forces) and Rayleigh (gravity). We will see further examples 
in other subsections.

In general, the velocity $U$ of a droplet in a concentration gradient is described with its so-called mobility
$M$, namely $U = M \nabla c$, with the mobility being determined by the droplet radius $R$, the concentration
dependence of the surface tension, and the dynamic viscosities $\eta_{i,o}$ of the 
inner and outer liquid, $M = R \partial_c \sigma / (2\eta_o + 3 \eta_i ) $ 
\cite{young1959,anderson1989,anderson1982jfm,izri2014},
provided no other phoretic forces play a role. 

Flow driven by concentration gradients and restricted to a few nanometers close to a {\it solid} interface is called 
{\it diffusiophoresis}  \cite{anderson1989,anderson1982jfm,marbach2017,prieve2019}. 
This term 
 in particular refers to the motion of a colloidal particle in a concentration gradient. 
Though both Marangoni flow and diffusiophoresis 
originate from the interplay of surfaces with compositional gradients, 
we will not focus on diffusiophoresis   in this review, as in general 
the term  refers to liquid-solid interfaces, and not to droplets. 
Note however that there are some exceptions; e.g.\ in refs.\ \cite{yang2018-stone,morozov2019} the movement of a charged 
 oil droplet in a solute gradient
is referred to as diffusiophoresis, as then a  discontinuity in the flow velocity across the droplet  interface arises, while 
the hydrodynamic stresses at the interface are continuous, see figure 1 of ref.\ \cite{morozov2019}.

\begin{figure}[htb]
  \centering
  \includegraphics[width=0.98\textwidth]{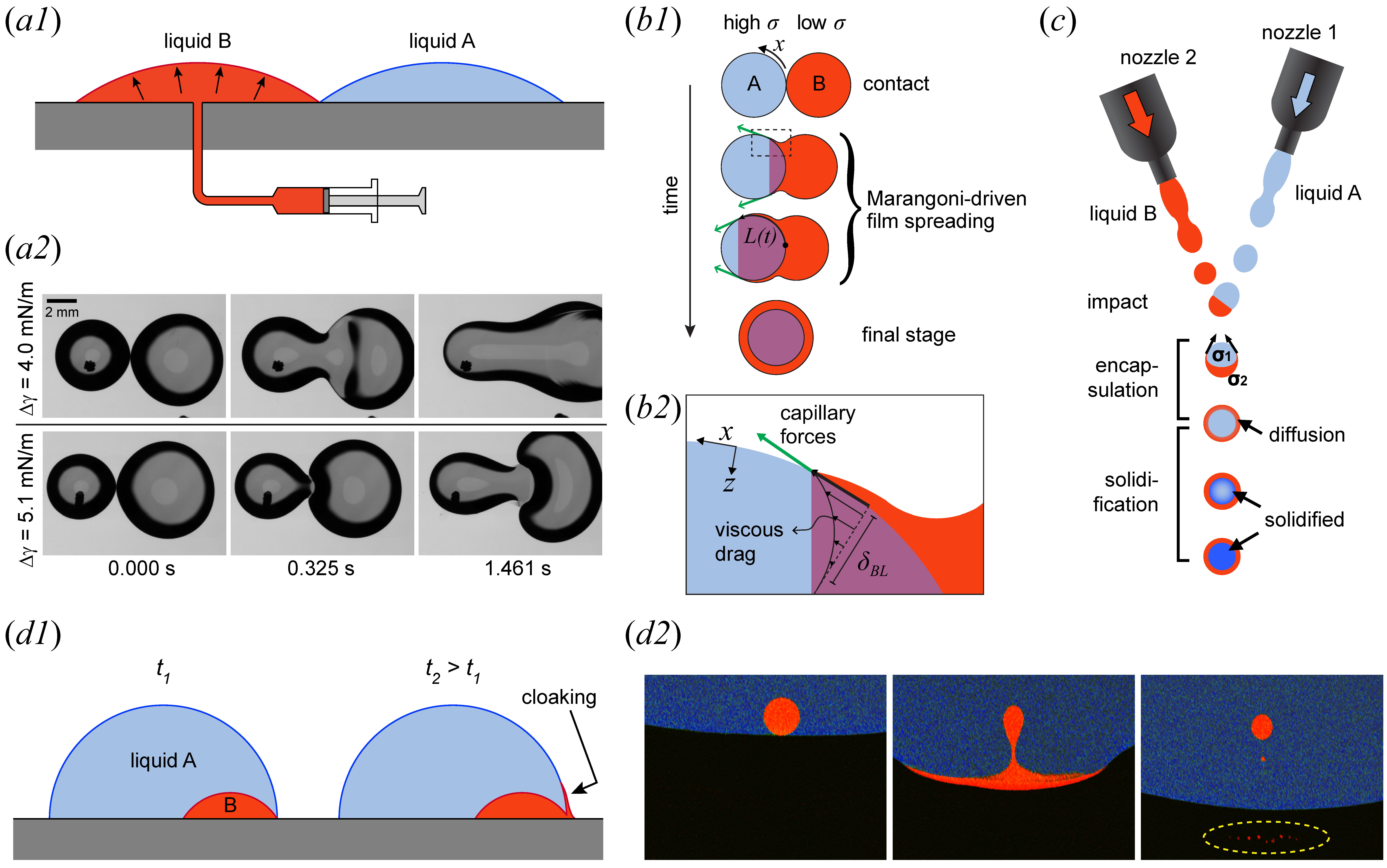}%
  \caption{\it
 Various types of coalescence of two droplets consisting of different liquids. 
 In (a1) a side-view sketch of a coalescent experiment of two {\em sessile} droplets is shown
  \cite{karpitschka2010,karpitschka2012col}.  The outcome is shown in  (a2), where 
  we  show  top-view snapshots  for two different surface tension differences $\Delta\sigma$ established
 with different 1,2-butanediol/water mixtures, 
 featuring  enhanced  (upper row) respective delayed (bottom row) coalescence. 
 Figures adapted from   \cite{karpitschka2010,karpitschka2012col}. 
 In (b1) we sketch a side view of two {\em pendent} droplets 
   \cite{koldeweij2019}, with the one with lower surface tension creeping over the one with higher surface tension.
   A detailed sketch of the process is shown in (b2), with the viscous velocity profile caused by the viscous force,  counteracting the Marangoni force,
 which here however wins, leading to encapsulation. 
   Sketches adapted from \cite{koldeweij2019}. 
 (c) Similarly, an  encapsulation is also possible  for {\em jetted} droplets of different liquids
 \cite{visser2018}. Here, one droplet  contains a cross-linker and the other a polymer, such that the coalescence leads 
  to solidification. 
  In (d1) we sketch two snapshots (at two different times $t_1$ and $t_2 > t_1$) of another geometry of droplet-droplet interaction, namely one sessile droplet sitting on another 
  smaller one.  When the two three-phase-contact-lines touch each other (left, $t_1$), part of the smaller droplet can be ``spit out'' (right, $t_2$).
  Three actual snapshots of confocal images at consecutive times are shown in (d2). Figures adopted from \cite{yu2019}.
   }
  \label{fig_coalesce}
\end{figure}

\subsection{Coalescence of droplets of different liquids} \label{coales}
A concentration gradient can instantaneously  be imposed by a collision of two droplets 
consisting of two different
liquids. In the case of 
coalescence of fully miscible droplets, the surface tension difference at the interface will lead to Marangoni flow,
with the droplet of smaller surface tension being pulled over that with the larger one, while the counteracting  force
in general is of viscous nature. 
The droplets can either be  sessile
\cite{riegler2008,karpitschka2010,karpitschka2012col} or  pendant \cite{koldeweij2019} from a nozzle with a large contact angle,  or even  hitting
each other in flight \cite{yeo2003,visser2018}, or -- both as sessile droplets -- sitting on top of each other \cite{yu2019},
see figure \ref{fig_coalesce}.

Note that the droplets need not be of equal 
volume and that the coalescence of a droplet with a pool of a different liquid is included
 in case (b) of figure \ref{fig_coalesce}. 
In that case, 
for large enough Ohnesorge and Marangoni numbers (i.e., dominance of Marangoni forces),  the spreading front $L(t)$ displays
a  universal scaling law $L(t) \sim t^{3/4} (\Delta\sigma)^{1/2} / (\rho \eta )^{1/4}$ 
or $L(t) / R \sim (Ma/Sc)^{1/2}  ~ (t \nu /R^2)^{3/4}$ (where $R$ is the droplet
radius)  over many orders  of magnitude \cite{berg2009,koldeweij2019,wodlei2018}. For case (a) of figure
\ref{fig_coalesce} 
-- sessile droplets -- the dynamics can be much
richer, as, depending on the surface tension difference $\Delta \sigma$, 
 it can either show enhanced coalescence similar to the case in figure \ref{fig_coalesce}b , or, more 
remarkably,  {\it delayed} 
coalescence, due to a competition between capillary pressure and dynamic pressure, induced by the Marangoni flow,
see figure \ref{fig_coalesce}a. 
   
The situation further complicates once the different liquids in the drops react with each other
\cite{jehannin2015} or a solidification (e.g.\ with cross-linkers) 
 is induced \cite{visser2018}, see sketch of figure \ref{fig_coalesce}c. Then the crucial new parameter is the 
Damk\"ohler number, which compares the time scale of the chemical reaction with that of the hydrodynamics. 
 Yet another situation occurs when one larger sessile droplet is immersing a smaller one of a different liquid, as shown in
 figure \ref{fig_coalesce}d. When the two three-phase-contact-lines touch each other, for certain combinations of surface tensions
 part of the interior sessile droplet can be ``spit out'' by the larger one \cite{yu2019}, see figure \ref{fig_coalesce}d.

\subsection{Droplets in concentration gradients emerging  from chemical reactions}\label{sec-swim}
In  the  examples of the two  previous subsections the concentration 
 gradient leading to Marangoni flow was imposed from outside. 
However, such a gradient can also evolve, e.g.\ thanks to chemical reactions. A famous example 
for such gradients can be achieved
with 
 so-called Janus particles, i.e., particles with different chemical composition on different sides  \cite{paxton2004}.   In the particular work of Paxton {\it et al.}\  \cite{paxton2004}   one side of the
Janus particle consisted of   gold and the other one  of platinum. 
When put in a $H_2O_2$ solution, on the Pt side catalytically generated oxygen nanobubbles emerged, leading to a chemical potential gradient, which induced a (generalized) Marangoni flow 
along the particle surface, pushing the particle backwards. There are 
many variations of such systems, employing the concentration gradient between 
front and rear, e.g.\ in refs.\ \cite{golestanian2005,jiang2010,buttinoni2013,michelin2014}. Note that only very small gradients $\nabla \sigma$
in surface tension are required to drive visible motion  of the microparticles \cite{maass2016}. 

For Janus particles, the symmetry-breaking is imposed on the particle. 
However, even more interestingly, such symmetry breaking leading to a gradient in concentration around
a particle or  droplet can also 
 emerge  {\it spontaneously},  due to an instability. 
Michelin {\it et al.} \cite{michelin2013} theoretically showed that for large enough Peclet numbers the combined 
effect of   solute   dissolution (solubilization) 
 (or  solute reaction)  at the droplet 
interface and Marangoni flow 
can produce an instability, resulting in 
spontaneous and self-sustained motion of the initially isotropic droplet \cite{golovin1989,rednikov1994,schmitt2013,izri2014,morozov2019b}. 
Here the Peclet number is defined
with the mobility $M$ of the droplet, the surface emission  rate $\alpha$ \cite{golestanian2005}, and the droplet radius $R$, namely
$Pe = M\alpha R^2 / D^2$. The crucial mechanism is that nonlinear mixing of the dissolved substance or the
reagent establishes gradients in concentration, which then are  converted into flow by Marangoni forces 
or diffusiophoresis   \cite{anderson1989,abecassis2008,palacci2010prl,banerjee2016pnas,banerjee2019}. 
The bifurcation diagram of this process and the resulting swimming velocities
are shown in figure \ref{young-diddens}e, along with the concentration fields around the droplets for various Peclet 
numbers, see insets of figure \ref{young-diddens}e. They and figure   \ref{young-diddens}a also give an idea 
on the flow field around the swimming droplet. 
 For even larger Peclet numbers, the droplet can curl \cite{krueger2016}, swim helically  \cite{suga2018} or even 
chaotically  \cite{izri2014,hu2019,morozov2019}; 
examples are
 shown in figure \ref{young-diddens}f,g. Once two or even more  of such droplets are close to each other,
they show very rich and highly nontrivial collective behavior \cite{thutupalli2011,palacci2013,moerman2017,lippera2020}.

  The solubilization or chemical reactions at the droplet interface can be of various nature. Many
   examples are
  given in the review article by Maass {\it et al.} \cite{maass2016}.
  These include simply dissolving or slowly reacting droplets, but also 
  micellar solubilization \cite{pena2006,herminghaus2014}
  of a
   surfactant on the oil droplet well above the critical micelle concentration (CMC), 
  nematic  
liquid crystal droplets self-propelling
 in a highly concentrated surfactant solution \cite{krueger2016,suga2018,hokmabad2019},
  droplets with a 
  surfactant undergoing a chemical reaction \cite{thutupalli2011}, 
  binary droplets
  with selective dissolution \cite{poesio2009},  and many others \cite{maass2016}. 
  Obviously, the driving strength  of solubilization or dissolution will decrease as function of time and the 
  droplet will shrink, leading to transitions in the motion pattern, e.g. from random, to helical, to straight 
  \cite{suga2018}, but this process is very slow and can take many hours.  
  An interesting case is also when the droplet crosses its own trace. 
  Another  flabbergasting case is a 
   surface-active  chemotactic droplet which is even able to navigate through a complex maze, thanks to an imposed 
  concentration gradient and the self-propelling property \cite{lagzi2010,cejkova2014,jin2017pnas}.
  Such an example is shown in figure \ref{young-diddens}d.

 

\begin{figure}[htb]
\begin{center}
\includegraphics[width=0.99\textwidth]{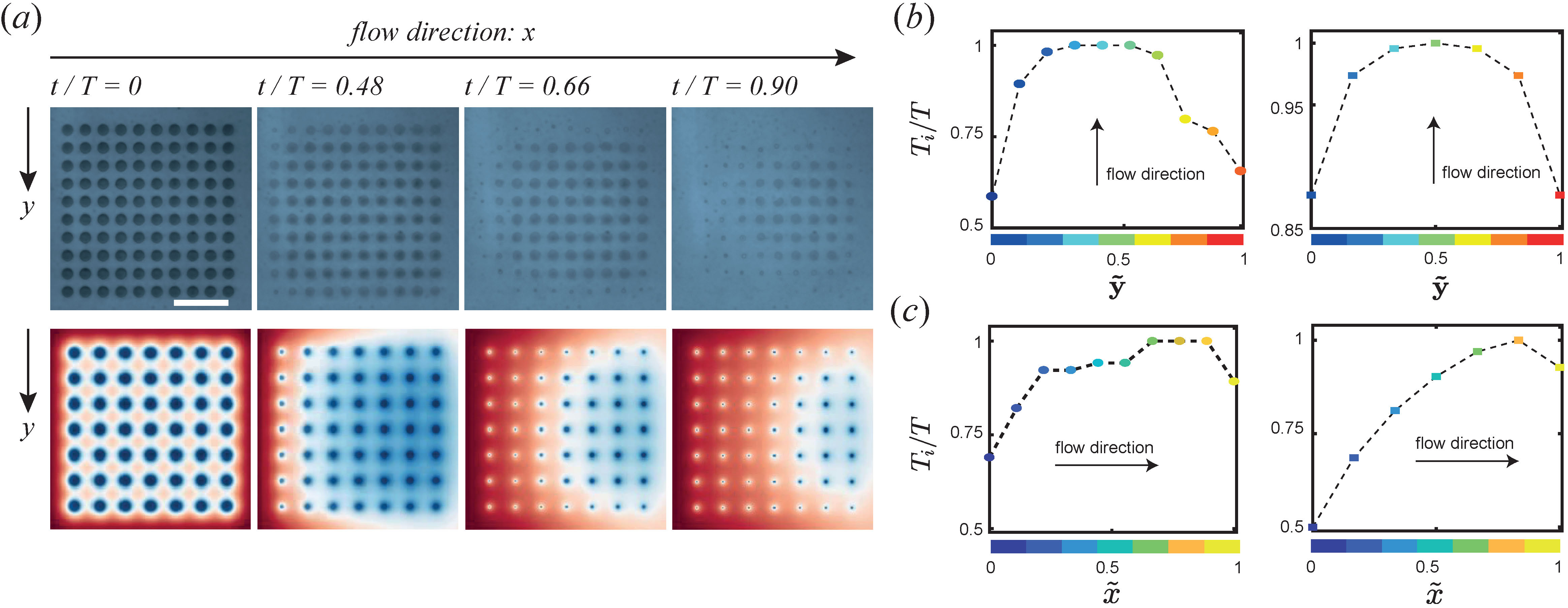} 
\caption{{  \it 
Collective dissolution for droplets: 
(a) Time evolution of droplet dissolution: 
The upper row shows the experimental results for a 10$\times$10 oil droplet array with a spacing of 5 $\mu$m for a flow rate of $ 200 \mu L/min$ 
(corresponding to a Peclet number of $Pe = 237 $) 
for four difference times $t$,  normalized using $T$, which is the total time for the droplet array to completely dissolve. 
The length of the scale bar is 45 $\mu$m.
The lower row shows a pseudocolour plot of the concentration field from numerical simulations for a 7$\times$7 droplet array for corresponding flow conditions. Blue and red corresponds to oil and water, respectively. The water is injected into the chamber along the 
 $ \hat e_\text x$ direction.  
(b) Total dissolution time ($T_i$) of droplets along the  span-wise direction y and (c) the stream-wise direction x. In (b) and (c), on the left the experimental data are shown, and on the right the numerical
ones. 
Figures adapted  from \cite{bao2018}. 
    }}
\label{fig-lei-bao}
\end{center}
\end{figure}

\subsection{Droplets in concentration gradients emerging  from phase transitions} \label{phase-trans} 
The simplest case  of a droplet in an out-of-equilibrium situation may be a volatile droplet evaporating in air or 
 a soluble droplet dissolving in another, originally pure  liquid. 
For a spherical droplet in the {\it bulk}  of a host liquid this problem was analytically solved 
by Epstein and Plesset \cite{epstein1950}. That  calculation was originally made for dissolving bubbles, but later generalized
to dissolving droplets \cite{duncan2006}. Also 
the evaporation or dissolution of a single {\it sessile}  droplet consisting of a pure liquid 
in a solvent 
has meanwhile reasonably well
been  understood \cite{picknett1977,deegan1997,hu2002,popov2005,cazabat2010,erbil2012,gelderblom2011,stauber2014,lohse2015rmp} -- even if it is not purely diffusive so that convective 
effects outside the droplet due to density differences of the liquids come into play \cite{dietrich2016a}, enhancing
the dissolution.

The situation becomes more interesting once {\it multiple}  dissolving or evaporating sessile droplets interact,
as in general they shield each other: The reason is that dissolving neighboring droplets 
 reduce  the concentration gradient at the interface and thus the outflux
from the droplet, leading to longer (and heterogeneous -- depending on the position of the droplet) life-times \cite{laghezza2016,carrier2016,zhu2018,michelin2018,bao2018,wray2020,chong2020}, even
 when the solvent is flowing over the sessile 
droplets \cite{bao2018}. Such a situation has recently been explored both experimentally and numerically, leading
to reasonable agreement, as seen in figure \ref{fig-lei-bao}. Note that the dissolution delay by shielding only holds
in the pure diffusive regime. Once convective effects come into play (i.e., density difference between the 
droplet and its host liquid with large resulting Rayleigh numbers), 
remarkably, collective effects can also {\it enhance}  dissolution,
due to the collective interaction of individual density plumes 
\cite{chong2020}.

\begin{figure}[htb]
\begin{center}
\includegraphics[width=0.95\textwidth,angle=-0]{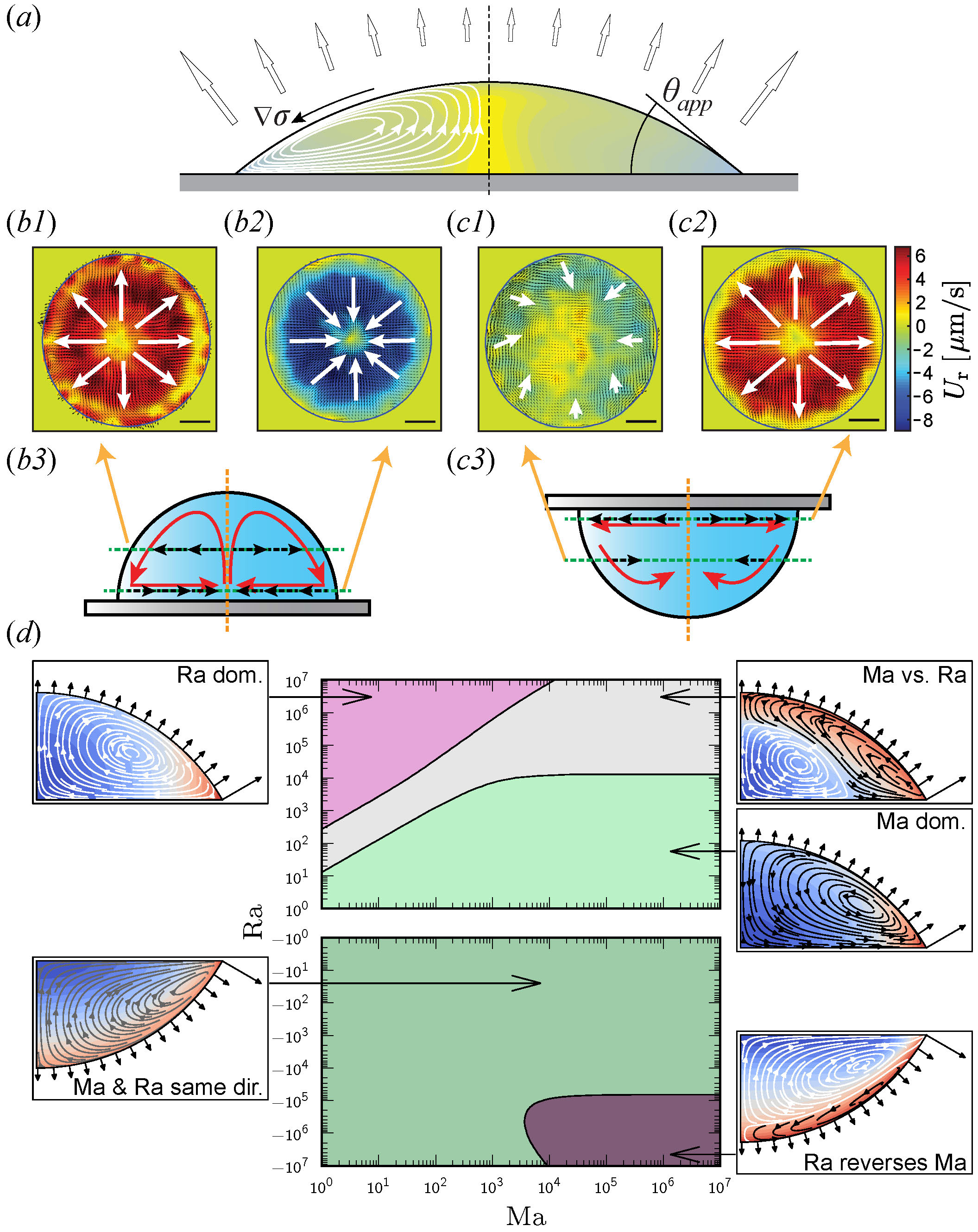} 
\caption{{  \it 
(a) 
Evaporation of a 5 pL 
binary droplet consisting of ethanol (0.1\%) 
and water. The higher evaporation rate at the rim
and the selective evaporation of the ethanol  leads to  a surface tension gradient $\nabla  \sigma$ at the droplet-air interface, driving
Marangoni convection inside the droplet. Note that for larger driving (i.e., larger Marangoni number), the axisymmetry of the process breaks. 
The evaporation of a binary sessile (b) and of a binary pendant (c) droplet: 
(b1/c1) and (b2/c2) show the measured velocity field within the sessile/pendant  droplet droplet at 
midheight/at the substrate,  as shown in the side view sketches (b3/c3). The different flow patterns in (b) and (c) 
clearly reveal that gravity (pointing downwards) plays a major role. 
Figure taken from \cite{li2019-yaxing}.
(d) Phase diagram in the $Ra$ vs $Ma$ parameter space for an evaporating binary 
 droplet: Depending on $Ra$ and $Ma$, either Rayleigh convection rolls are seen, or Marangoni convections
  rolls, or both. We also show the respective flow and concentration  patterns.
 Figure taken from \cite{diddens2020}.
    }}
\label{mega2}
\end{center}
\end{figure}

Coming back to single sessile droplets, 
 concentration gradients can also emerge from selective evaporation (or dissolution) from a sessile {\it binary}  
droplet (or -- again more generally -- a droplet consisting of several  miscible liquids). The reason lies in a combination
of the (in general) different volatility  of the components and the singularity of the evaporation (or dissolution) rate
at the rim of the droplet \cite{deegan1997,deegan2000,popov2005,cazabat2010} (provided the contact angle is smaller
than $90^o$), leading to a concentration gradient at the interface of the droplet. The resulting
surface tension  gradient drives a Marangoni flow inside the
droplet 
\cite{scriven1960,hu2005,bennacer2014,tan2016,kim2016,dietrich2017,karpitschka2017,diddens2017,kim2017-stone,li2018-yaxing,kim2018,edwards2018,li2019-yaxing,marin2019}, see figure \ref{mega2}a.  
This effect is very similar to what leads to 
  the so-called tears of wine \cite{hosoi2001} inside a partially filled 
  wine-glass: In this  case,  selective evaporation of ethanol at the edge of the meniscus 
  leads to larger surface 
  tension, which thus pulls part of the remaining wine upwards along the alcohol-wetting 
   glass, finally
  leading to an instability of the film and droplets sinking  down the  glass wall. 
  In evaporating droplets, the Marangoni convection can  be so violent -- with velocity exceeding mm/s -- that the 
axisymmetry breaks  \cite{diddens2017,kim2016}. 
The same can happen through moisture absorption: Shin {\it et al.} \cite{shin2016} show that the absorption of water vapor (i.e., moisture) to 
a sessile or pendent   glycerol droplet can lead to an axisymmetry breaking 
 B\'enard-Marangoni instability of the resulting flow, which is driven by water concentration gradients at the
droplet-air interface, emerging from preferential absorption of the water at the rim of the droplet. 
The evaporation, absorption, and dissolution dynamics 
of multicomponent sessile droplets presently is a very active area of research.

Depending on the nature of the two components  of the binary liquid, the (stronger) evaporation of one component  can lead to several different scenarios
for the physicochemical hydrodynamics of the sessile binary droplet. 
In a first scenario \cite{li2018-yaxing,kim2018,karpitschka2018}, the selective evaporation of one liquid can lead to segregation
of this liquid in the center of the droplet thanks to shielding by the other, non-volatile liquid. This scenario occurs once
the Marangoni flow arising through concentration differences  at  the droplet-air interface
 is too weak to fully mix the two liquids.

An even more interesting second scenario comes into play once  the two miscible liquids of the binary droplet 
display a sufficiently large density gradient \cite{edwards2018,li2019-yaxing}, as can become 
 the case for e.g.\ 
a droplet consisting of water and glycerol \cite{li2019-yaxing}. This density gradient is ``activated'' by selective evaporation of 
the more volatile liquid, which in ref.\ \cite{li2019-yaxing} is water. In 
dimensionless form, it is expressed as Rayleigh 
number $Ra$. The gravitational forces resulting from the density gradient compete with the Marangoni forces on the 
droplet-air interface, whose strength in dimensionless form is expressed as Marangoni number $Ma$. 
Once the gravitational forces win, Rayleigh convection rolls dominate and the flow pattern in sessile droplets 
(figure \ref{mega2}b) is  very 
different from that in pendant droplets (figure \ref{mega2}c). For sub-millimeter droplets this is remarkable as the
Bond number of such droplets is $Bo \ll 1$, on first sight implying that 
 gravity  does not play a role -- but only as compared to 
capillarity (implying the spherical-cap-shape of the droplet), but not as compared to viscosity. 
The full phase diagram in the $Ra$ vs $Ma$ parameter space is displayed in figure \ref{mega2}d, featuring  either pure
Rayleigh convection rolls, or pure Marangoni convection rolls, or both at the same time. Also for large Rayleigh numbers
the axisymmetry can be broken, and a Rayleigh-Taylor instability can emerge \cite{li2020-yaxing}.

\begin{figure}[htb]
  \centering
 \includegraphics[width=0.96\textwidth]{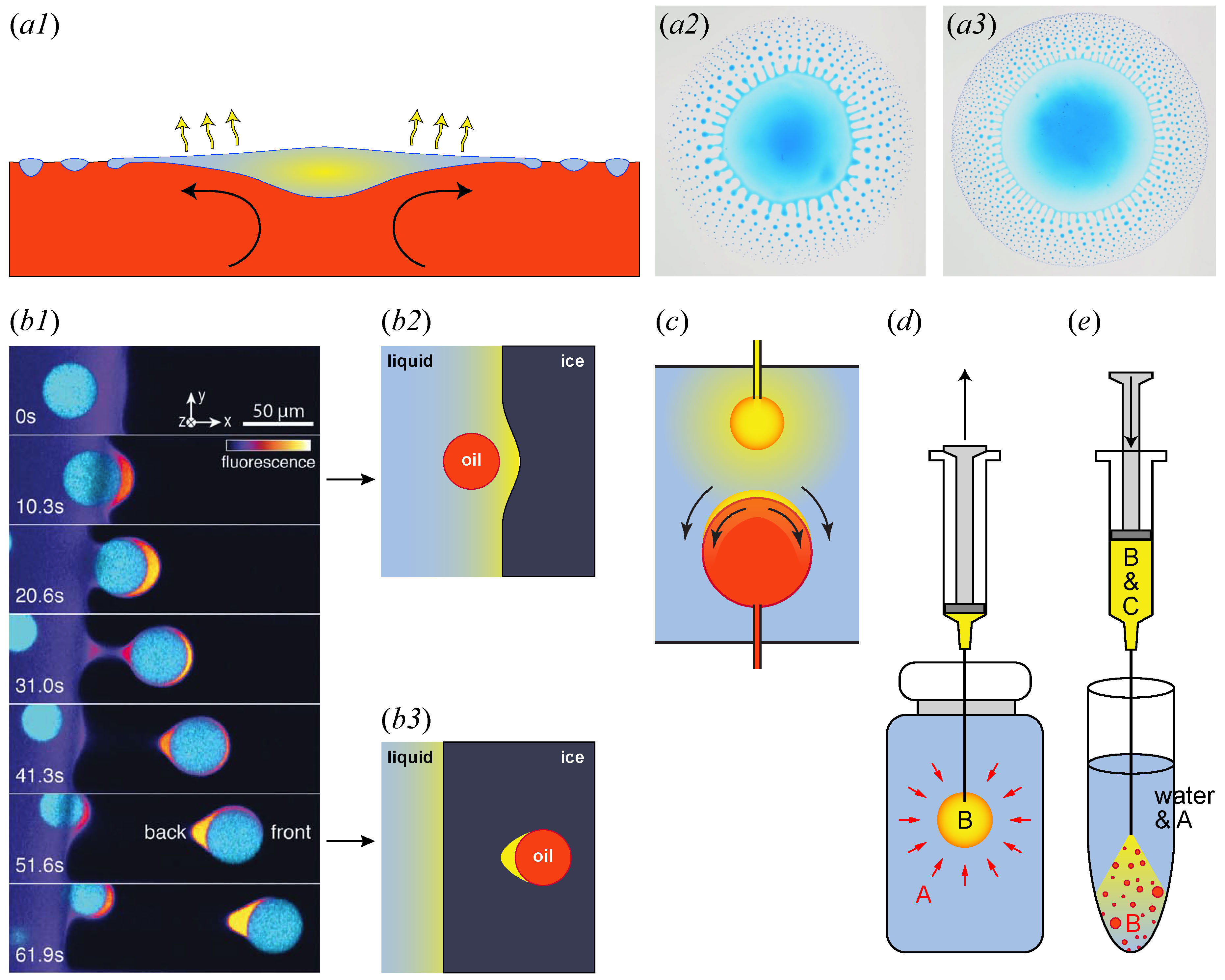}
  \caption{\it (a) 
 Marangoni bursting: Evaporation-induced emulsification of a two-component droplet (isopropanol and water)
 on an immiscible bath. 
 (a1) Sketch of the mechanism. (a2) and (a3) Top view of the flow pattern for two different isopropanol  concentrations 
 of the deposited binary alcohol-water droplet, from lower ((a2), 0.4 mass fraction) to higher ((a3),  0.45 mass fraction).
 Figures taken from \cite{durey2018} (after adoption). 
 (b1) 
Typical interaction between a droplet and the solidification front, showing the accumulation of solute and its redistribution around the droplet. Horizontal cross section. Here the  temperature gradient leading to the freezing is  5 K/mm, with a 
sample velocity of $2\mu m/s$. 
Figure taken from \cite{dedovets2018}.
The sketches (b2) (early times) and (b3) (later times when the front has passed over the oil droplet) visualize the process. 
(c) Two droplets of two different slightly soluble liquid are approached to each other in a third liquid and interact diffusively,
  even before they coalesce. 
(d) Classical single droplet microextraction, as developed by Pregl \cite{pregl1917}.
(e) Principle of Dispersed Liquid-Liquid Microextraction (DLLME), as first suggested in ref.\ \cite{rezaee2006a}.
For an explanation we refer to the text. 
  }
   \label{mega3}
\end{figure}

Instabilities can also emerge once a binary droplet evaporates on a bath of an immiscible liquid  
\cite{keiser2017,durey2018,wodlei2018}, see figure \ref{mega3}a. Here, similarly as in subsection \ref{coales}, 
the Marangoni stress in the drop pulls the droplet outwards but is balanced by the viscous stress in the oil bath, 
leading to a bulk flow (black  arrows in figure \ref{mega3}a1).  
The evaporation causes a transition from partial to complete wetting, which sets the radius of 
the central drop in figures \ref{mega3}a2,a3. 
In the limiting case of the bath height going to 
zero, one has the situation of a binary droplet 
 on a thin film. Here the outwards Marangoni flow can get so 
violent that a hole emerges in the film \cite{hernandez2015}.

The dynamics of an evaporating binary sessile droplet -- be it on a substrate or a liquid -- is further
complicated once also surfactants come into play
\cite{kim2016,kim2017-stone}. Also in this case the Marangoni driving of the flow can be so strong that its 
axisymmetry is 
broken. 
The surfactants
 and the composition of the binary droplet also strongly affect the pattern of the deposit 
as Kim {\it et al.} demonstrated with evaporating whisky droplets \cite{kim2016}.

Note that not only the dissolution of sessile binary  droplets with the resulting Marangoni flows is highly non-trivial,
as elaborated above, but even the dissolution of spherical binary droplets in the bulk: The reason is that, 
due to  selective dissolution,  
concentration gradients inside the droplet between its surface and its bulk can arise \cite{maheshwari2017}. These can lead to 
considerable memory effects (in particular for small droplets), which
 do not exist for the dissolution of a pure droplet  \cite{chu2016p,lohse2016ff,maheshwari2017}.

Another example for a  concentration gradient (i.e., an  out-of-equilibrium situation) 
which emerges from phase transitions is 
 a freezing miscible mixture  of different  liquids with different freezing temperatures. 
A particularly intriguing case 
is a pure immiscible droplet in a binary liquid undergoing solidification of one component \cite{deville2017}. 
Then a concentration gradient is emerging, acting on the droplet, either pushing it away from the solidification front
or towards it, possibly finally leading to the engulfment of  the droplet 
\cite{dedovets2018}. An example is shown in figure \ref{mega3}b.
A generalization of this problem are freezing colloids which show extremely rich and intriguing behavior
 \cite{deville2017}.

\subsection{Droplets in  ternary liquids: Nucleation  and growth}\label{ouzo} 
\begin{figure}[htb]
\begin{center}
\includegraphics[width=0.98\textwidth]{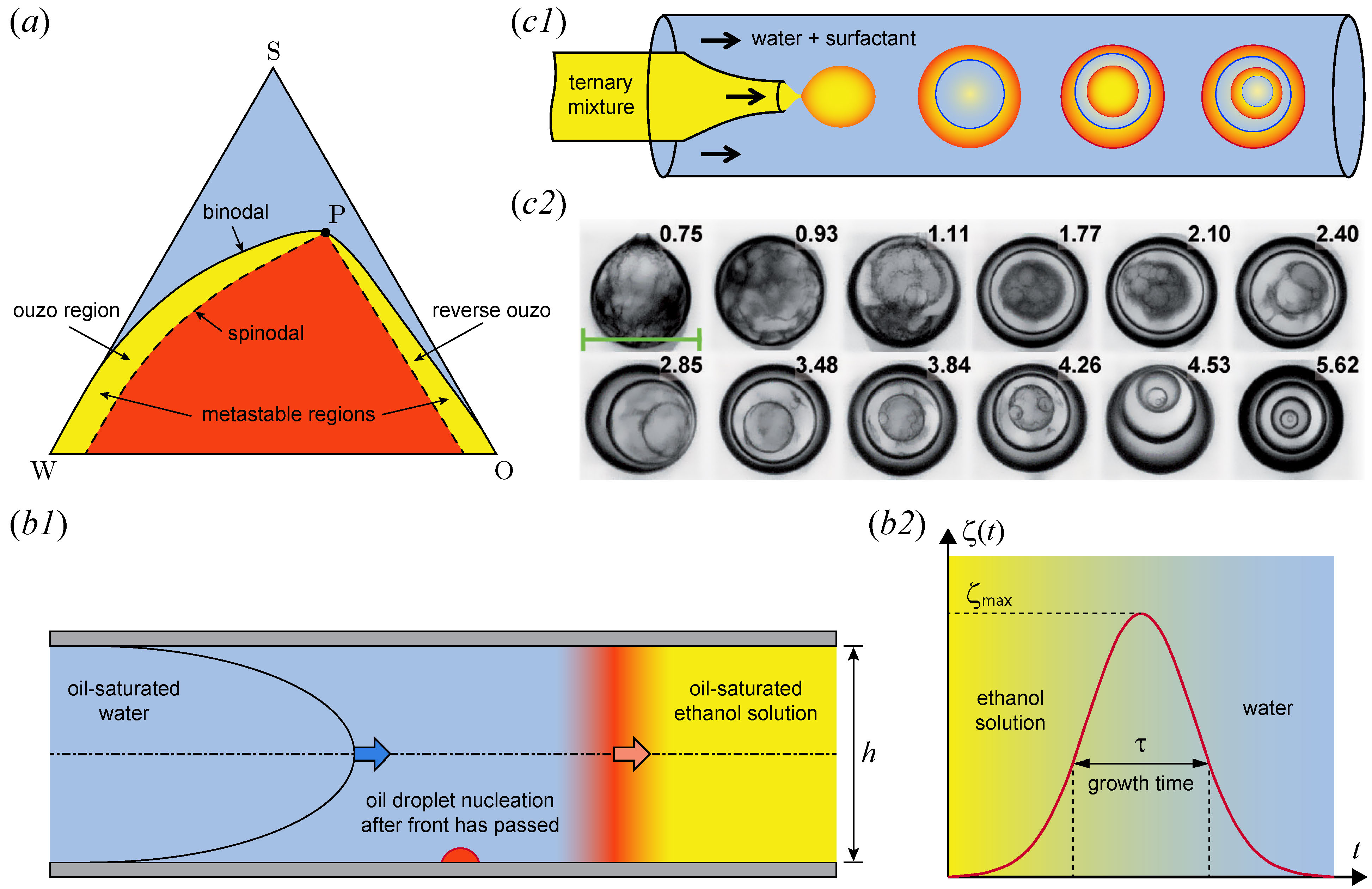} 
\caption{{  \it 
(a) Schematic ternary diagram of a ternary mixture of water (w), a solute (S) like ethanol, and an oil (O)
like anise oil in the case of ouzo. At the  corners, the respective liquid has a 100\% concentration. The concentration then linearly decreases to zero at the other two corners, along the axes of the triangle. 
Above the binodal curve, the three liquids are fully miscible (blue region). Below
the spinodal curve (red region), water and oil separate into two phases. In between the binodal curve and the spinodal curve, there is a small region
(yellow)  in which sub-micrometer-sized oil droplets
 in water (``ouzo region") or water droplets in oil (``reverse ouzo") are metastable. Figure 
adapted  from \cite{solans2016}. 
More complicated ternary diagrams with so-called pre-ouzo regimes with oil nanodroplets in an otherwise miscible regime
are possible \cite{lopian2016}. 
Principle of solvent exchange:
(b1) A bad solvent (here water) is replacing a good solvent (here ethanol), leading to a front of oversaturation and thus 
to the nucleation of (here oil) droplets at the (hydrophobic) surface. Note that the plug-flow like nature of the oversaturation
front is due to Taylor-Aris dispersion \cite{taylor1953dispersion,aris1956dispersion,aris1959dispersion,yu2017}. 
The velocity profile itself remains parabolic.
(b2) For fixed position at the wall, a front of oversaturation $\zeta(t) >0$ is passing by, leading to droplet growth.
Figure  adapted  from \cite{zhang2015pnas}.
(c1) Ternary liquid of  oil, water, and ethanol 
 brought into contact with an aqueous phase with a co-flow device \cite{anna2003,utada2005}, leading to phase-separation in 
 the emerging droplets, which develop onion-like structures. 
 (c2) Details of the developing onion-like structures of a phase-separating ternary liquid of (a). The time is given in seconds. 
 The length-scale given in the first snapshot at $0.75s$  is $100\mu m$. Figure taken from \cite{haase2014}. 
    }}
\label{mega4}
\end{center}
\end{figure}

The examples up to now dealt with miscible,  sparingly miscible, and immiscible liquids, but the solubility 
of one liquid in another one  was always constant. However, in ternary liquids, this need not be the case: The mutual 
solubilities in general depend on the relative concentrations of the liquid. This is commonly expressed in the so-called
ternary diagram, see figure \ref{mega4}a for a sketch of a  typical situation. In particular, the ternary diagram can contain 
so-called ouzo-regions, in which sub-micrometer-sized droplets of one species can metastably exists. The 
best-known example for such liquids is indeed ouzo itself.

Ouzo is a transparent 
Greek liquor (equivalently, one can take as example the French 
Pastis, the Italian Sambuca,  or the Turkish Raki), 
which chemically -- in spite of its actual  more complex composition --  for our purposes can be considered 
as  a ternary liquid consisting of ethanol, water, and (anise) oil. 
When served, water is usually added, which lowers the solubility of the oil (see the ternary diagram 
of figure  \ref{mega4}a), thus leading to 
oil oversaturation and subsequently to  the nucleation of 
oil droplets in the bulk liquid and thus to the 
characteristic milky color. This process is called ``ouzo effect''  \cite{vitale2003,ganachaud2005,lepeltier2014} or also 
 {\it solvent exchange} or {\it solvent shifting}. 
The ouzo emulsion is amazingly stable 
against Ostwald ripening \cite{voorhees1985,solans2005}, i.e., the capillary pressure driven shrinkage of 
the smaller droplets and the simultaneous growth of the larger ones (coarsening). 
 Moreover, the nucleated droplets have 
 a relatively sharp size distribution. Both features -- the absence Ostwald ripening and the sharp droplet size distribution --
 are not fully understood and an active area of research 
 \cite{zemb2016}.

When the  solvent exchange  process
 takes place in the 
presence of a hydrophobic surface, {\it sessile}  nanodroplets will nucleate at this 
 surface
and then grow 
 \cite{zhang2012softmatter,lohse2015rmp}. 
 This offers the opportunity for a {\it bottom-up approach}  in ``building'' droplets or also  crystals. 
 Nucleation and growth  of the droplets strongly depend on the geometrical
and chemical nature of the substrate and  pinning of the contact line  respective the absence thereof plays a paramount role 
\cite{joanny1984,gennes1985} in determining the growth mode of the droplet (constant contact radius (CR) mode
vs constant contact angle (CA) mode). It is this pinning that enables the stability of surface nanodroplets and nanobubbles
against evaporation or dissolution  \cite{liu2013pre,lohse2015,lohse2015rmp}.

The advances in modern microfluidics and microscopy of various kinds (see box 2) allowed to monitor 
the growth of the nucleated sessile nanodroplets as function of the control parameters 
\cite{zhang2015pnas,yu2015,yu2017,zeng2019} -- including on patterned surfaces \cite{bao2016,peng2016-coll},
as function of time \cite{dyett2018growth}, and including 
collective effects of the nucleating droplets \cite{peng2015acsnano,xu2017,dyett2018-1}. 

The essence  of the process is sketched in figure \ref{mega4}b: A bad (but fully saturated with a solute) 
solvent in a narrow channel (of submillimeter height $h$) 
is replacing
 a good
one  (also saturated or at least containing sufficient solute), 
leading to a front of oversaturation of the solute, which passes by the substrate. This passage
leads  to nucleation of nanodroplets on the substrate (provided the wetability of the droplet liquid on the 
 substrate allows for this) 
and to their growth. This growth is controlled by the thickness of the diffusive boundary layer around the droplets
(which is set by the Prandtl-Blasius-Pohlhausen boundary layer theory \cite{schlichting1979})
and thus the mean flow velocity, or, in dimensionless numbers, the Peclet number $Pe$. One can theoretically derive
$\left< Vol_f\right> \sim h^3 Pe^{3/4}$ for the final area averaged volume $\left< Vol_f\right> $ of the droplets 
\cite{zhang2015,yu2017}, which agrees very well with the experimental data.

\begin{figure}[htb]
\begin{center}  
\includegraphics[width=0.99\textwidth,angle=-0]{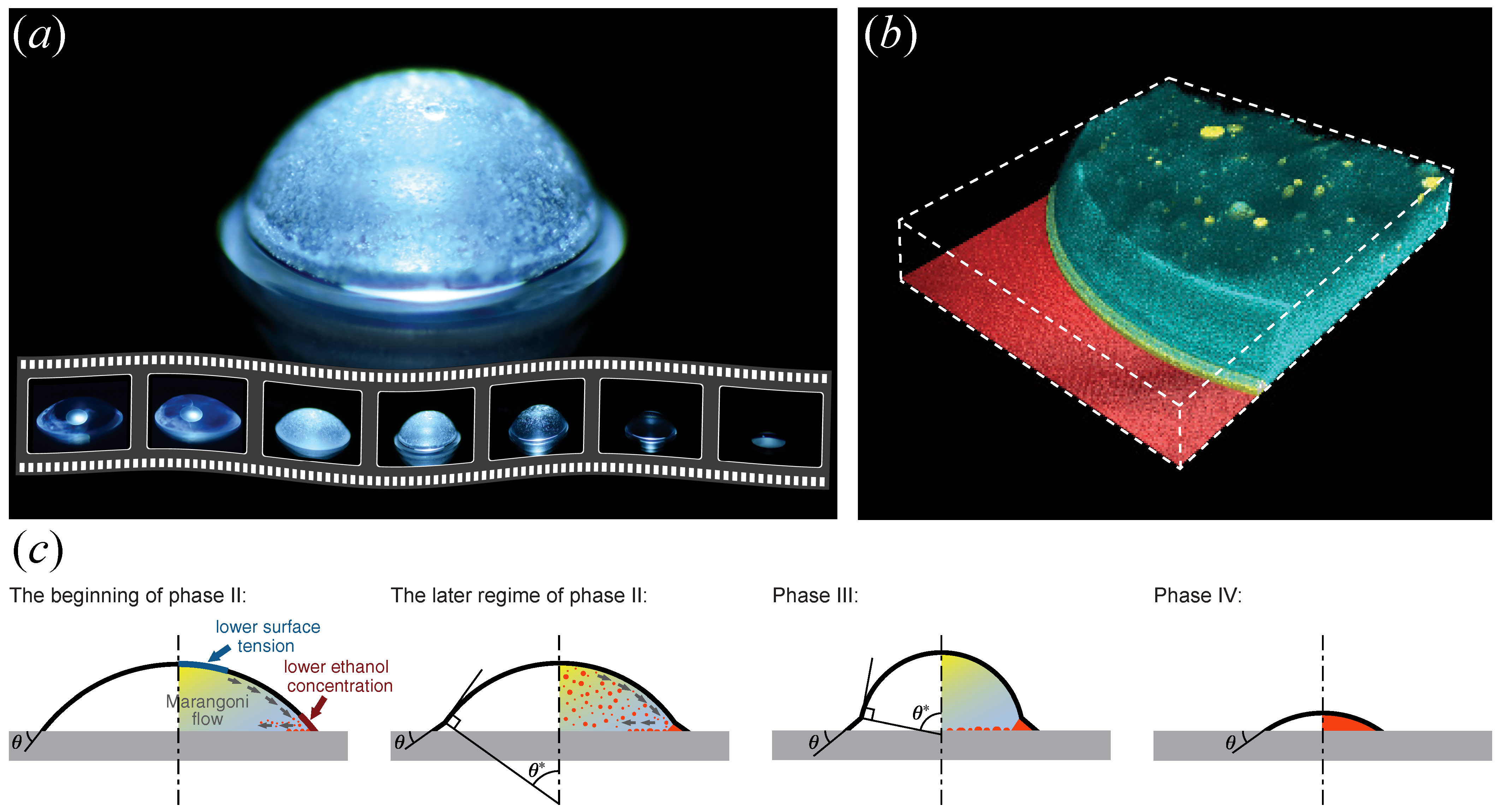} 
\caption{{  \it 
Evaporating ouzo droplet: 
(a) Optical image of phase III of the droplet, with 
a milky droplet sitting on an oil ring. The seven insets show earlier and later snapshots of the droplet.
(b) Confocal image early in phase III, when the oil ring starts to develop. 
(c) The four sketches show different  lifetime phases of the   evaporating ouzo droplet, see the text for more details. 
Figures adapted  from \cite{tan2016}.
    }}
\label{fig-ouzo}
\end{center}
\end{figure}

The solvent shifting 
can also be driven  by  evaporation or dissolution  of one of the solvents. 
Such experiments combine features of the Marangoni flow in ternary droplets 
triggered by  evaporation or dissolution as described in subsection \ref{phase-trans} with the nucleation of microdroplets 
triggered by solvent shifting as described in this subsection. 

A very illustrative and simple  example for such an experiment is the evaporation of an ouzo droplet on a substrate under ambient conditions \cite{tan2016,diddens2017},
see 
figure \ref{fig-ouzo}, where we show optical and confocal snapshots and sketches of the 
 four different stages of 
the evaporation process 
of the ouzo droplet  \cite{tan2016,diddens2017}: 
In phase I, the spherical cap-shaped droplet remains transparent, while the more volatile ethanol is evaporating, 
preferentially at the rim of the drop due to the geometric 
singularity there, as explained in subsection \ref{phase-trans}. 
This leads to a local ethanol concentration reduction and
according to the ternary diagram figure \ref{mega4}a to oil droplet nucleation 
at the rim. This is the beginning of phase II, in which oil microdroplets quickly nucleate in the whole drop, leading to 
the typical  milky ouzo appearance. These microdroplets can coalesce and form an oil ring at the rim of the droplet
 (early in phase III). 
At some point  all ethanol has evaporated and  the drop, which now has a characteristic non-spherical-cap shape
with the water drop sitting on top of the oil ring, 
thus is  transparent  again (late in phase III). Finally, in phase IV, also all water has evaporated, leaving behind a tiny spherical cap-shaped oil drop. The entire evaporation process
 takes about a quarter of an hour. 
Note that this  example of an evaporating ouzo droplet on a substrate can also numerically be 
treated, employing finite element methods, resulting in a very good agreement between numerics and experiments
\cite{tan2016,diddens2017}. 

There are various variations of the theme of an evaporating or dissolving ternary droplet: In ref.\ \cite{tan2017}
an  ouzo droplet evaporating on a superamphiphobic surface is studied, again both experimentally and numerically. 
In this case the contact angle is much larger than $90^o$, resulting in a maximal evaporation rate at the apex of the droplet
and correspondingly to a start of the oil microdroplet nucleation at that position. Obviously, the Marangoni flow 
is then also towards the apex of the droplet, and not towards the rim, as it was the case for contact angles smaller than
$90^o$. In ref.\ \cite{tan2019} the dissolution of a water-ethanol drop in a bath of anise oil was analyzed: Here during the dissolution, two types of microdroplet nucleation was observed, 
namely oil microdroplet nucleation in the aqueous drop (``ouzo effect") and 
water microdroplet nucleation in the surrounding oil (``inverse ouzo effect"), see 
figure \ref{mega4}a. Again, various
physicochemical hydrodynamical processes such as Marangoni flow, Rayleigh convection, diffusion and nucleation compete
in an extremely rich way, but can nonetheless be disentangled \cite{tan2019} by  including
 the key physics  in a simple model
\cite{kirkaldy1963,ruschak1972,miller2007,tan2019}.  

An even richer system with two big drops  consisting of different aqueous liquids 
(one water drop, and one consisting of ethanol-water mixtures)  in a bath of toluene as host
liquid 
 is studied in ref.\ \cite{otero2018}. The diffusion of the aqueous liquids through the toluene leads to the nucleation
 of aqueous microdroplets (inverse ouzo effect) in between the two bigger drops. 
A variation of this experiment is sketched in figure  \ref{mega4}c1: Two droplets of two different slightly soluble liquids
 are approached to each other in a third liquid. Even before they touch (a situation we had discribed in subsection \ref{coales}), there will be diffusive exchange of matter. 
 In figure  \ref{mega3}c the solubility of the yellow liquid is larger than that of the red one, leading to a concentration gradient of the yellow liquid on the red drop and thus in general to Marangoni forces and the resulting Marangoni flow in the red drop and in between the drops. This will dramatically change the diffusive process and the forces between the droplets. In case of ternary liquids with a solubility gap it can also lead to nucleation of microdroplets, as in ref.\ \cite{otero2018}.

Droplets of ternary liquids  can phase separate in many  different ways 
\cite{choi2013,haase2014,zarzar2015,lopian2016,zemb2016,lu2017,moerman2018}, showing extremely rich and complex behavior.  For example, a miscible droplet of diethylphthalate (DEP) oil, water, and ethanol,
which has a ternary diagram similar to that shown in the sketch of figure \ref{mega4}a,  
demixes upon contact with an  aqueous phase,  to give alternating, onion-like  layers of oil and water \cite{haase2014,moerman2018},
see figure \ref{mega4}c.     The detailed  dynamics  of this liquid-liquid phase separation and 
the type of emerging structures   strongly depends on the exact relative 
initial composition of the 
ternary liquid and their viscosity. 
The process is controlled by diffusion of water into the ternary droplet and vice versa of its components out
of the droplet. The liquid-liquid phase separation  can be modelled with Cahn-Hilliard type approaches  \cite{moerman2018}. 
        
A microfluidics co-flow device similar to that of figure \ref{mega4}c1 operated with ternary liquids is also very well suited to impose a well-defined concentration gradient,
in order to quantitatively study  the competition between diffusion, Marangoni convection, and nucleation of microdroplets
in the ouzo-region or inverse  ouzo-regime of the ternary diagram
 \cite{hajian2015}. 
 The reason is that, with the controlled and laminar flow profiles of such a device, 
 detailed experimental information on  the conditions for droplet formation and their radial migration in the ternary flow 
 can be obtained and compared with the ternary phase diagrams.

\section{Relevance and applications of physicochemical hydrodynamics in droplet  systems}\label{relevance}
 
 After having shown some  examples of recent fundamental work on the 
 physicochemical hydrodynamics in well-defined droplet systems in the
 previous 
  section, we will now come to the relevance of multicomponent droplet systems in applications. 
 In the introduction we have already  mentioned 
 various application fields. In the following  subsections  we want to go into more 
 detail for five quite different application fields. Also for these applications we cannot be encyclopedic, nor go into depth,
 but we hope to convey the flavor of the applications and their great potential. 
 As in the previous  section 
 we ended with ternary droplets with a solubility gap, here we will directly continue with them,
 and only then come to miscible binary and multicomponent droplets.

\subsection{Chemical analysis and diagnostics}

 Liquid-liquid extraction -- the transfer of a solute from one solvent to another --
 is one of the core processes in chemical technology and analysis. 
 For chemical analysis such as chromatography, 
ever since the pioneering work of the Nobel Laureate  Fritz Pregl
\cite{pregl1917} on microanalysis, there have been continuous efforts 
to further miniaturize the extraction process of the analyte and to optimize the extraction recovery and  preconcentration factor.
The driving factors for the miniaturization have continued to increase in recent years \cite{jain2011}, reflecting the
urgency of the problem: First, the need to detect 
{\it trace quantities} of some substance is still increasing  in the medical, biomedical, 
food safety,  and environmental  context. 
Next, the health monitoring systems of the future will be based on {\it rapid} measurements on {\it small} sample volumes. 
Finally, the miniaturization will  lead to less chemical waste and environmental strain, i.e., towards ``greener''
analytical methods  and process technologies. 

In the last two decades so-called {\it single-drop microextraction}  \cite{jain2011} 
has become very popular for sample preparation 
of trace organic and inorganic analysis. The principle of this method is shown in 
 figure \ref{mega3}d. Here a solute A dissolved in water accumulates in a droplet of water-immiscible liquid B, due 
 to 
 its higher solubility in B as compared to water. After an equilibrium has been achieved, the droplet, which now consists of a mixture of A and B, is extracted with a
  syringe,  to be further analysed by e.g.\ chromatography. 
 
 The scale on which single-drop microextraction can be done obviously
  remains limited, but this limitation is overcome in the modern technique 
of 
{\it dispersive liquid-liquid microextraction} (DLLME), which was invented less than 15 years  ago 
\cite{rezaee2006a,rezaee2010,zgola2011} and heavily makes use of the solubility gap of ternary liquids and thus 
the ouzo effect as explained in subsection \ref{ouzo}. 
Here, a mixture of two miscible liquids B and C (with low concentration of B) 
is put into water containing  the analyte A, with B being immiscible with water (say, carbon-tetrachloride \cite{zgola2011}), 
but C being miscible (say, acetone \cite{zgola2011}), see figure  \ref{mega3}e. 
When poured into the aqueous solution containing A, droplets of B will
immediately nucleate and then further grow out of the oversaturated solution.  
The liquid B is chosen such that it  has much higher solubility for the analyte A than water 
 and  is  heavier than water and liquid C. The large total surface of the 
 nucleated microdroplet ensemble will greatly help the extraction process.  
The final step is to centrifuge the dispersion and take out the A-B phase. 

The relevant parameters to characterize the performance of 
liquid-liquid extraction processes are the so-called preconcentration factor, defined as the ratio 
of the analyte 
concentration in the centrifuged droplets, and the so-called extraction recovery, defined as the percentage of 
analyte which could be extracted.
Hitherto, it has not been possible to {\it a-priori calculate} 
  these parameters, hindering the optimalization of liquid-liquid extraction processes, which presently is 
often done by trial-and-error. Even liquid-handling robot systems \cite{parrilla2014} 
were built to automate the evolutionary process
of finding the optimum. Clearly, a quantitative understanding of the nucleation and the 
diffusive dynamics of droplets in ternary systems is crucial to make progress towards a quantitative understanding
and systematic optimization of DDLME. 

 Another diagnostic application of the physicochemical hydrodynamics of droplets far from equilibrium is 
 nanoextraction
 of tracers \cite{ocana2016}. Sample preparation is considered to be the most difficult step in analytic workflow. Current methods for extraction and separation of minute amounts of 
 substances in liquid samples are laborious, time-consuming, often involve large amounts of toxic organic solvents, and are  difficult to automatize, implying high costs of man-power. 
 However, liquid-liquid extraction and online analysis of traces of analytes in aqueous solutions, including in biomedical, health, pharmaceutical and environmental contexts, may be dramatically improved by  surface nanodroplet-based sensing
  techniques \cite{li2019functional,li2019automated,qian2020one}.
  The basis of such  nanoextraction are
  surface nanodroplets pre-formed by solvent exchange on a substrate within a microflow channel. The principle is that the partition coefficient of the compound in the droplets is much higher than in the solution. The compound in the liquid 
  can thus be extracted into the surface nanodroplets and be quantified by surface-sensitive spectroscopic techniques
  \cite{qian2020one}. 
  This  approach can potentially achieve extraction-detection of analytes at extremely low concentrations in one single and simple step, allowing for fully automized sample analysis through programmable nanodroplet production for extraction and in-situ detection.

 \subsection{Pharmaceutics and chemical \& environmental  engineering}

The ouzo effect 
 is  not only relevant for microextraction in chemical analysis and diagnosis, but also 
for drug production and delivery \cite{lepeltier2014} and in pharmaceutical science
 and cosmetics  \cite{gutierrez2008},  providing a  basis for the preparation of 
 pharmaceutical products, formulation of cosmetics
and insecticides, liquid-liquid extraction and for many other practical processes. 
In this context the ouzo effect is  also called ``nano-precipitation''. 
Small
hydrophobic organic molecules, lipids, or polymers also exhibit similar microphase separation and, 
by mixing them with a poor solvent, 
form nanodroplets  or nanoparticles with homogeneous sizes.  
Note that often the details of the mixing process matter.  Why this is the case 
 has  not yet been fully understood. E.g., in some case the 
mixing must be very rapid -- a process then called 
``flash nanoprecipitation'', which has been demonstrated to be a simple way to produce drug-loaded polymeric 
nanoparticles, protein nanoparticles, and other multifunctional colloids with narrow size distribution 
 \cite{zhang2012sm,akbulut2009,zeng2019ahm}. 
Droplet formation by solvent exchange is also highly relevant for separation and purification in other 
 applications of the chemical and pharmaceutical industry, including undesirable oiling-out crystallization in production of pharmaceutical ingredients, amino acids, and proteins \cite{gerald2014}. 

Also on a larger scale,
 understanding and controlling nucleating  and growing 
 droplets in liquid-liquid phase separation is essential for improving the efficiency of industrial operations,
 e.g.\  in extraction of  natural resources, recycling of valuables from waste and removing organics in waste water 
 treatment, in flotation, and in renewable energy technologies.

Finally, dilution-induced phase separation (i.e., the ouzo effect) 
 is  important for  advanced  
oil recovery processes  \cite{sun2017application}. 
An  example 
 is paraffinic froth treatment in the industrial process of oil sands extraction \cite{qliu2013}. 
Heavy oil bitumen with a
considerable amount of fine solids and water is separated from oil sands ore by warm-water
extraction to form bitumen froth. Solids and water are
removed by adding sufficient light alkane (which is a poor solvent for asphaltenes, a solubility family of 
extremely heavy species of bitumen) to achieve
 a critical solvent/bitumen ratio. 
Asphaltene precipitation triggered by this dilution then forms agglomeration with water drops and
solids, sweeping the solids and water off under
gravity and producing diluted bitumen of high
quality which is easy to transport \cite{he2015interfacial}. 
Control of the size distribution and morphology of asphaltene precipitates may lead to a
 more efficient oil recovery approach, with less hydrocarbon loss to  waste \cite{sun2017application}.

 \subsection{Synthetic chemistry and biology}
  
 Apart from encapsulation and precursors for nanoparticles as described in the previous subsection, the 
 droplets and their physicochemical hydrodynamics
  also find many applications in chemical synthesis. They provide a confined environment in synthetic chemistry to realize cascade reactions in a fashion similar to artificial cells. The surface of microdroplets serves as a biphasic site
for catalysts to access immiscible liquid phases inside and outside of droplets, significantly
improving the
specificity and efficiency of interfacial catalytic reactions for biofuel upgrading \cite{crossley2010}. 
The droplets generated from chemical reactions were found to preferentially and differentially segregate and compartmentalize RNA, suggesting that droplets may play a
 role in protecting essential chemical components for the  origin of life  \cite{jia2019}.  
 Indeed, the droplets can grow and divide from addition of materials produced in droplet reaction, resembling 
 prebiotic protocells \cite{zwicker2014,zwicker2017,golestanian2017}.

Microdroplets can also act as microlabs or chemical microreactors,  and often chemical
reactions, which do not take place in bulk water, do occur in microdroplets \cite{lee2015,lee2019}. 
To account for the enhanced chemical reactions in the microdroplets, 
a  reaction-adsorption mechanism was proposed   \cite{fallah2014}. According to this mechanism, the molecules at the interface are more active, which is attributed to the solvation energy 
getting available thanks to solubilization. The energy required for the molecules at the interface to react is therefore less than that in the bulk reaction \cite{fallah2014}. 


As an entity of microcompartment, droplets formed from aqueous liquid-liquid phase separation of polyelectrolytes have drawn intensive interest from biology and cell research.  
The process of 
  phase separation of an aqueous solution consisting of two oppositely charged polyelectrolytes into two immiscible aqueous phases 
  --  coacervate droplets and supernatant -- 
is called ``coacervation'' 
   \cite{overbeek1957}.  
 The dense coacervate droplets are rich in polymers while the supernatant is a dilute equilibrium phase, poor in polymers. Coacervation has a long history of utilization  in
encapsulation applications \cite{kizilay2011}. 
The process of coacervation  was also  proposed  as crucial in  theories  of abiogenesis (the development of life) 
\cite{oparin1957}.
Coacervation is driven by both entropy and electrostatic interactions between polyelectrolytes, influenced by molecule weight, charge distribution and chirality of polymers, the concentration and the
weight ratio of the two polymers and the  ionic strength of the aqueous medium. 
Prior to the macroscopic phase separation, the size distribution of the polyelectrolyte complexes becomes very narrow. Intermolecular complexes of several hundred nanometers form until macroscopic phase separation occurs.

The research interest in this subject has recently further intensified
 due to the functional roles of intracellular coacervates and their relation to diseases \cite{shin2017}, 
 and due to the extraordinary properties of coacervates as new materials,  namely 
 underwater superglue, deep tissue bonding or bone fixation, scaffold coatings, bone cement, and 
  drug encapsulating materials \cite{seo2015}. 
What remains largely unknown is the evolution of nanoscale coacervates.  Model polyelectrolytes with defined molecular weight and charge distribution have been studied to understand the effect of charge patterns \cite{chang2017}. 
It would be interesting to quantify the 
 temporal evolution of the coacervate nanocolloids during 
  the phase separation process, to better understand its 
   fluid dynamics.

\subsection{Inkjet printing} 
Binary and multicomponent droplets obviously are extremely relevant in inkjet printing \cite{wijshoff2010,kuang2014,hoath2016,dijksman2019}
as nearly all inks are multicomponent
and not only contain pigments and surfactants, but also consist of various 
different  liquids, with different surface tensions,
volatilities, and viscosities. Moreover, different nozzles of an
 inkjet printer are operated with different inks. Both features imply  that 
selective evaporation and coalescence of droplets of different liquids as described in subsections \ref{phase-trans} and \ref{coales} 
are very crucial processes in inkjet printing.

This not only holds for droplets on the substrate towards which the droplets are jetted but also on the nozzle-plate of the
inkjet channel, close to the nozzle out of which the droplets are jetted, and to the meniscus of the ink itself. Here selective 
evaporation of one component of the ink can be a major problem, as we will demonstrate with two 
 examples from 
piezoacoustic inkjet printing \cite{wijshoff2010}, which is one of the most advanced and most controlled forms of inkjet printing: 
\begin{itemize}
\item 
In piezoacoustic inkjet printing, the ink in the nozzle is well mixed, thanks to the acoustic field driving the jetting of the droplets.
Therefore the material properties of the ink such as surface tension or viscosity 
are determined by  the  composition  of the various liquid components. Now 
 imagine 
a nozzle 
at rest for some printing cycles, as that particular ink is not needed. In that time -- say many seconds -- one or more 
components of the ink can selectively evaporate out of the nozzle. 
This changes the material properties of the remaining ink such as the surface
tension and therefore the 
required  pulse strength for jetting once the nozzle is activated again  -- a major source of inaccuracy. 
\item 
 Even worse, droplets of ink on the nozzle plate can selectively
evaporate, introducing a concentration gradient and thus surface tension gradient on the nozzle, leading to a Marangoni flow
\cite{jong2007,beulen2007}. Once this flow is directed towards the nozzle it can lead to transport of dirt particle into the nozzle which 
can lead to air bubble entrainment and nozzle failure \cite{jong2006b,fraters2019} -- a major disaster for the printing process. 
\end{itemize} 

Also on the substrate on which the droplets are jetted  Marangoni flow within one droplet \cite{scriven1960,tan2016,kim2016,dietrich2017,karpitschka2017,diddens2017,kim2017-stone,li2018-yaxing,kim2018,edwards2018,li2019-yaxing,marin2019}  (see figures \ref{mega2}a,b,c)
or in between different droplets \cite{karpitschka2010}  (figure \ref{fig_coalesce}) can lead to unwanted effects, in particular as these flows transport pigments. 
Or does the emerging Marangoni flow between the droplets perhaps even  help in mixing the droplets (``bleeding of mixed colors'')? 
Yet another question is how the multicomponent nature of the evaporating droplet and the resulting flows affect
the coating pattern \cite{kim2016,sefiane2014,kuang2014,han2012,cai2008}. 
The jetted multicomponent droplets on the substrate may in addition undergo
 phase changes by evaporation,   solidification, or  chemical reactions. 
In the latter case of 
{\it chemical reactions}  inside the droplets, 
including reactions leading to {\it solidification}, as e.g. with crosslinkers \cite{visser2018}, many questions arise: 
What does exactly happen when one reactant diffuses into a droplet of another 
reactant, e.g.\ from a neighboring droplet or from some reservoir?
How does the reaction propagate from the droplet surface where the two reactants first make contact?
How does a two-component paint ``chemically dry'' through cross-linkers? 
``Watching paint dry''  is in fact  interesting, relevant, and 
largely unexplored science! It is also very 
timely, because of sharpened  environmental regulations with respect to the evaporation of (toxic) solvents.

\subsection{Nanotechnology and nano- \& micro-materials} 
The solidification process after drop-drop coalescence -- with one droplet filled with crosslinkers and the other one
with a liquid responsive to it -- has indeed been used to manufacture micro-particles of controlled shape and size 
at very high  rate
\cite{visser2018}. As the process takes
 place after collision of such droplets in air (see figure \ref{fig_coalesce}e), 
Visser {\it et al.} \cite{visser2018} 
called it  
  ``in-air microfluidics''. 
 Its  control parameters are  -- next to the droplet sizes and their velocities -- the
  droplet compositions, through  which different degrees of Marangoni-flow-driven encapsulation can be achieved,
  leading to controlled 
  mono-disperse emulsions, particles, and fibers with diameters of 20 to 300 $\mu m$. 
  
  Also ternary liquids have been used to manufacture micro- and even nanoparticles, namely by employing the 
  ouzo effect -- or then also called nano-precipitation. In fact, with this process micro and nanomaterials can be built  bottom-up  in a 
  well-controlled way \cite{reverchon2007,chan2011,huang2018}.

Another important material science application for which the 
physicochemical hydrodynamics of droplets is relevant  are 
freezing colloids \cite{deville2017}, as e.g.\ shown in figure \ref{mega3}b. 
Here the crucial question is: How do immersed droplets interact with the freezing front in binary or multicomponent freezing liquids? 
On the one hand particle-reinforced metal alloys require homogeneous distribution of particles in the matrix, 
and an immediate engulfment is therefore preferred. 
On the other hand,  for single-crystal growth, a complete rejection of impurities is obviously crucial \cite{dedovets2018,zhang2006steel}.
A rejection is also preferred whenever engulfed droplets or bubbles would introduce uncontrolled defects 
into a cooling  alloy.

Finally, a very important application of the physicochemical hydrodynamics of droplets is in the
semiconductor industry and in nanotechnology where 
surfaces  like wafers have to be extremely clean and dry. This can be achieved by the so-called 
Marangoni drying \cite{leenaars1990,marra1991,obrien1993,thess1999,matar2001,karpitschka2017}, where gradients in surface tension
are imposed  to drive the liquid of a thin film outwards, e.g.\ through local vapor deposition 
of a different fluid (often isopropyl alcohol -- IPA), 
which absorbs on the film,  or through deposition of a multicomponent droplet \cite{hernandez2015}. 
For Marangoni drying to work properly, the interaction of sessile droplets on the wafer is of course essential 
and unexpected hinderance of coalescence of droplets of different liquids as shown in refs.\ \cite{karpitschka2010,karpitschka2012col} (see figure \ref{fig_coalesce}a2) are obviously a major problem. 

 In the context of drying wafers with their nanoscale structures,
  capillary forces can also cause major problems \cite{okorn2014}: When a sessile 
   droplet attached to different structures on a surface  is evaporating, the capillary forces, which are tremendous
   on the tiny length scales relevant on wafers, pull  these structures together, which can make them break 
   \cite{bico2004,duprat2012,bico2018}.

 \section{General Lessons}\label{conclusions} 
We hope to have given an idea  
how in the last years the community is working towards a deepened 
 quantitative   understanding of  
physicochemical hydrodynamics of droplets far from equilibrium, in order to 
illuminate the fundamental fluid dynamics  of 
immersed (multicomponent) (surface) droplets. 
 To do so we have shown 
 controlled experiments and numerical simulations
for idealized setups, allowing for a  one-to-one comparison between experiments and numerics/theory, 
in order to test the theoretical understanding.

Much remains to be done, both from a fundamental point of view, but also from the application side, with many new 
and extremely relevant applications of the physicochemical hydrodynamics of droplets far from equilibrium popping up 
at a very high rate. But we expect that the rapid progress in this field is continuing, as from our point of view we are living 
in the golden age of fluid dynamics: The reasons are
the continuous increase in 
 computational power, so that simulations which we did not dare to dream of even ten years ago are now possible and that  
 a similar revolution (for the same reason) is taking place  in digital
high-speed imaging, thanks to which we can now routinely resolve the millisecond time scale and even smaller scales, revealing new physics on these scales, which up to now was inaccessible, and producing a huge amount of data on the flow. 
Moreover, other advanced equipment like confocal microscopy, digital holographic microscopy and atomic force microscopy are coming to be used more and more in fluid dynamics. Considering all of these advances together, the gap between what can be measured and what can be simulated {\it ab initio} 
 is narrowing more quickly than we had anticipated at the end of the last century.
Other gaps are also closing. Fluid dynamics is bridging out into various neighboring disciplines, such as chemistry, and in particular colloid science, chemical analysis and diagnostics, lab-on-a-chip mircrofluidics, 
catalysis, electrolysis, medicine, biology, computational and data science, among many others. 
Here the techniques, approaches and traditions from fluid dynamics can offer a great deal of help to solve outstanding problems. Vice versa, these fields can offer wonderful questions to fluid dynamics.

Finally, anyone who dismissed the experiments with tears of wine or with coffee, whisky or ouzo drops  as a gimmick is very 
much mistaken: On the one hand, as we have seen, we learn important physics, chemistry, fluid dynamics, and 
colloid and 
 materials science from these systems and 
 processes, with many modern applications in diagnostics, pharmaceutics, biology, medicine, 
 chemical and environmental engineering, inkjet printing, nanotechnology, and micro-manufacturing, 
 all of huge relevance.  
 On the other hand -- and here we come to the education in science which  may be  even more  important 
 than one or the other application --
  the deciphering of these everyday phenomena are examples {\it par excellence}
   of the physicist's approach: first translate an observed phenomenon into a clean experiment with well-defined control parameters, then make precise observations and record data, then develop a theory and a model and confirm these by calculations and numerical simulations, and finally make predictions about how the system will behave with values for the control parameters from an even wider range. 
  Studying the fluid physics of these everyday systems is therefore also very suitable for the conceptual training of 
   PhD students, who can learn in this way, simultaneously and by example, clean experimentation, theory formulation, modelling and numerical simulation, which 
   from our point of view is  much more motivating and broader than being a small cog for a specific detail of a large 
   large-scale experiment with thousands of scientists involved.

\section*{Acknowledgements} 
We thank our colleagues, postdocs, Ph.D.\  students, and students 
 for all their work and contributions to our understanding of physicochemical hydrodynamics of droplets
  and for the stimulation and intellectual pleasure we have enjoyed when doing physics together. 
  In the context of the subjects covered in this review we  would like to name 
  Lei Bao, 
  Kai Leong Chong, 
  Pallav Kant, 
  Ziyang Lu, 
  Andrea Prosperetti, 
  Vamsi Spandan, 
  Michel Versluis,
  Claas-Willem Visser, 
  Herman Wijshoff, 
  Haitao Yu, and in particular
   Christian Diddens,
   Yanshen Li,
  Yaxing Li, and Huanshu Tan.
  Moreover, we thank Anne Juel, Stefan Karpitschka, 
  Corinna Maas, Andrea Prosperetti, Jacco Snoeijer, and 
  Howard Stone for comments on the manuscript. 
D.L.\  also acknowledges  Dennis van Gils for drawing many  figures 
and 
support from NWO under several projects, 
from the ERC-Advanced Grant ``DDD'' under the project number 740479,  and from the 
ERC-Proof-of-Concept Grant ``NanoEX'' under the project number 862032. 
X.H.Z. acknowledges support from the Natural Science and Engineering Council of Canada (NSERC) and from
the Canada Research Chairs program. 

\vspace{1cm}



\end{document}